\documentclass[aps,pre,twocolumn,titlepage,nofootinbib]{revtex4-2}
\usepackage[utf8]{inputenc}
\usepackage[T1]{fontenc}
\usepackage{graphicx}
\usepackage{color}
\usepackage{dcolumn}
\usepackage{latexsym}
\usepackage[normalem]{ulem}
\usepackage{hyperref,amssymb}
\usepackage{url}
\usepackage{physics}
\usepackage{tensor}
\usepackage{color,soul}
\usepackage{graphicx}
\usepackage{verbatim}
\usepackage{multirow}
\usepackage{amsmath}
\newcommand{\beq}{\begin{eqnarray}}
\newcommand{\eeq}{\end{eqnarray}}
\usepackage{mathrsfs}
\usepackage{float,soul}
\usepackage{mathtools}
\usepackage{slashed}
\usepackage{physics}	
\usepackage{graphicx}   
\usepackage{epstopdf}
\usepackage{tabularx}
\usepackage{hyperref}   
\usepackage{bbold}
\hypersetup{colorlinks,
}
\usepackage{wasysym}
\usepackage{feynmp}
\usepackage{hyperref}
\hypersetup{colorlinks,
}

\newcommand{\new}[1]{#1}

\begin{document}
\title{Relaxed hydrodynamic theory of electrically driven non-equilibrium steady states}
\author{Daniel K.~Brattan$^{1,2}$}
\author{Masataka Matsumoto$^{3,4,5}$}
\author{Matteo Baggioli$^{3,4,5}$}
\email{b.matteo@sjtu.edu.cn}
\author{Andrea Amoretti$^{1,2}$}
\email{andrea.amoretti@ge.infn.it}
\affiliation{\textbf{1} Dipartimento di Fisica, Università di Genova,
via Dodecaneso 33, I-16146, Genova, Italy}
\affiliation{\textbf{2} I.N.F.N. - Sezione di Genova, via Dodecaneso 33, I-16146, Genova, Italy}
\affiliation{\textbf{3} School of Physics and Astronomy, Shanghai Jiao Tong University, Shanghai 200240, China}
\affiliation{\textbf{4} Wilczek Quantum Center, School of Physics and Astronomy, Shanghai Jiao Tong University, Shanghai 200240, China}
\affiliation{\textbf{5} Shanghai Research Center for Quantum Sciences, Shanghai 201315, China}

\begin{abstract}
\noindent The capability of hydrodynamics to accurately describe slow and long-wavelength fluctuations around non-equilibrium steady states (NESS), characterized by a stationary flow of energy or matter in the presence of a driving force, remains an open question. In this study, we explicitly construct a hydrodynamic description of electrically driven non-equilibrium charged steady states \new{in the limit in which the relaxation of the first non-hydrodynamic excitation is parametrically slow}. Our approach involves introducing gapped modes and extending the effective description into a relaxed hydrodynamic theory (RHT). Leveraging the gauge-gravity duality as a tool for controlled computations within non-equilibrium systems, we establish an ultraviolet complete model for these NESS that confirms the validity of our RHT. In summary, our findings provide a concrete realization of the validity of hydrodynamics beyond thermal equilibrium, offering valuable insights into the dynamics of non-equilibrium systems.
\end{abstract}

\maketitle

\section{Introduction} 
 Hydrodynamics, in its broadest sense, offers a framework for understanding the long-wavelength and late-time dynamics of systems near thermal equilibrium, that extends beyond the realm of fluids to encompass diverse phenomena like electron flow in materials \cite{Varnavides2023}, spin wave dynamics in magnets \cite{PhysRev.188.898}, and vibrations in solids \cite{PhysRevB.13.500,PhysRevA.6.2401}. \new{To apply hydrodynamics to non-equilibrium steady states (NESS), which emerge as an external force propels a system, requires substantial generalizations (\textit{e.g.}, \cite{RevModPhys.32.25,PhysRevA.23.1451,PhysRevA.19.1290,PhysRevLett.42.287,Law1989}).}

\new{This issue is part of a bigger problem, namely the development of a statistical mechanic description for systems out of equilibrium. In fact, non-equilibrium steady states violate some of the axiomatic principles of statistical physics, \textit{e.g.} detailed balance, and their fluctuations and large deviation functions cannot be rationalized using the  physical concepts valid for equilibrium states. Several methods have been developed in the recent years to tackle this problem, such as the matrix ansatz, the additivity principle or macroscopic fluctuation theory \cite{derrida2007non}. In particular, for driven diffusive systems, macroscopic fluctuation theory (MFT) \cite{PhysRevLett.87.040601,Bertini2002} (see \cite{Bertini_2015} for a review) provides a unified macroscopic treatment of such states. Nevertheless, a macroscopic and hydrodynamic description of general NESS is still a subject of active investigation and several open questions remain. In our work, we will contribute to this program by constructing a concrete hydrodynamic theory for \textit{certain} (to be defined in detail later on) NESS.}

To illustrate the emergence of a NESS and subsequently the need to modifying standard hydrodynamics in such a situation, let us delve into the theory of heat conduction. Historically, this is described by the Cattaneo equation \cite{maxwell1867iv,cattaneo1958forme}, $ \left(1+\tau \partial_t\right)\,\vec{\mathcal{Q}}=-\kappa \vec{\nabla}T$, which links the thermal flux $\vec{\mathcal{Q}}$ to temperature gradients by means of the thermal conductivity $\kappa$. We notice that the ``\textit{thermal inertia}" parameter $\tau$ plays a fundamental role in this description since it makes the Cattaneo equation compatible with causality \cite{RevModPhys.61.41}, providing an early example of what is now defined as a quasi-hydrodynamic theory \cite{Grozdanov:2018fic,Jain:2023obu}. \new{The new parameter $\tau$ signifies the time delay required to establish a consistent heat conduction state within a volume element once a temperature gradient is applied. Its introduction solves the \textit{paradox of heat conduction}, namely the fact that Fourier's law predicts that thermal signals propagate with infinite speed. Intriguingly, in the case of most metals, $\tau$ is remarkably brief, of the order of picoseconds. This is the rationale for its common omission. Nevertheless, it is worth noting, with a hint of curiosity, that in sand, $\tau$ extends to approximately 21 seconds \cite{chandrasekharaiah1998hyperbolic}!}

Despite its wide applicability, the Cattaneo heat equation faced challenges in describing systems in motion with a constant background velocity $\vec{v}$ \cite{PhysRevLett.94.154301}. This issue was resolved by the Cattaneo-Christov model \cite{CHRISTOV2009481,doi:10.1098/rspa.2014.0845,STRAUGHAN201095}, by incorporating a material derivative that restores Galilean invariance,
$\left[1+\tau \left( \partial_t+\vec{v}\cdot \vec{\nabla}\right)\right]\,\vec{\mathcal{Q}}=-\kappa \vec{\nabla}T$. This model now finds applications in very diverse systems, from turbulent flows \cite{doi:10.1098/rspa.1995.0124} to chemotaxis \cite{Berezovskaya_1999} and traffic flow \cite{JORDAN2005220}.

Upon these modifications, the equation for heat conduction takes a more general form,
\begin{equation}\label{eq:cattaneo}
  \tau  \frac{\partial^2 T}{\partial t^2}+\frac{\partial T}{\partial t}+ \vec{v} \cdot \frac{\partial \vec{\nabla} T}{\partial t} - D \nabla^2 T = \mathcal{R} ,
\end{equation}
with $D$ being the thermal diffusivity and $\mathcal{R}$ covering possible external forces. Setting $\tau=\vec{v}=0$, one recovers the standard heat diffusion equation. 

In recent years, the ability of hydrodynamics to elucidate the intricate flows of electrons and charge conduction in solids \cite{Varnavides2023} has acquired particular significance owing to the engineering of ultra-clean 2D materials, like graphene \cite{Lucas_2018}, and the proliferation of strongly coupled materials, exemplified by high temperature superconductors \cite{Tranquada_2020}, for which a comprehensive microscopic theory remains elusive.

Within solid materials, a steady state can be intuitively achieved through the application of an external electric field. As illustrated by the conventional Drude model \cite{chaikin2000principles}, conduction electrons subjected to a constant external electric field exhibit a constant drift velocity, wherein the accelerating force is counteracted by the scattering mechanisms that deteriorate the average electronic momentum. The \new{dynamics of the} fluctuations surrounding this steady state becomes of crucial importance, not only for our comprehension of fundamental conduction properties \cite{10.1063/1.5143271} but also to establish a model of transport in strongly coupled materials where charge is not carried by well-defined quasiparticles \cite{doi:10.1126/science.abq6100}.

Constructing a robust hydrodynamic theory for a charged fluid under a constant background electric field poses formidable challenges. Early attempts at seamlessly integrating an electric field into hydrodynamic theory presuppose complete compensation by the variation of the chemical potential, resulting in an unphysical equilibrium configuration with a null drift velocity \cite{Kovtun:2016lfw} \new{(see nevertheless MFT \cite{Bertini_2015} for an alternative approach)}.

\new{
In particular, working with systems of infinite spatial extent, it is a result of translational symmetry that such systems will have an infinite DC conductivity; thus we must break this symmetry to have a finite DC conductivity. Consequently, in our case, dissipation effects must necessarily be incorporated into the theory to counterbalance the  external  driving electric field. Notably, both the electric field and the dissipation rates will become } ``thermodynamic variables'' within the system, treated on equal footing. Furthermore, the persistence of a constant background drift velocity unavoidably breaks boost invariance, necessitating the application of a boost-agnostic hydrodynamic formalism \cite{deBoer:2020xlc} to achieve a comprehensive understanding \cite{Amoretti:2022ovc,Amoretti:2024jig}. 

In what follows we develop the quasihydrodynamic theory around such a NESS \new{in the limit in which the relaxation rate of the spatial electric current is parametrically slow}. To assess the reliability of our hydrodynamic theory, we compare our predictions against a strongly coupled microscopic UV complete model, the probe brane setup \cite{Karch:2002sh}, achieved by means of the gauge-gravity duality. \new{We emphasize that the computations performed in the microscopic model do not rely on any hydrodynamic approximation. In fact, one of the advantages of using the gauge-gravity duality is the possibility of describing the complete out-of-equilibrium dynamics, beyond the late-time and large distance regime and beyond the approximation of considering only conserved (or almost conserved) quantities.}

\section{Relaxed hydrodynamic theory}
We consider a three dimensional system with uniform charge density $\rho$ and a stationary charge current satisfying $\vec{J}=\sigma_{\mathrm{DC}}(\rho,\vec{E}^2) \vec{E}$, where $\sigma_{\mathrm{DC}}$ is the nonlinear DC charge conductivity. \new{For simplicity, we will work in the \textit{probe limit} in which the stress tensor and its fluctuations decouple and we will ignore them in what follows. This approximation is valid when the number of degrees of freedom of the bath is parametrically larger than the number of degrees of freedom of the system coupled to it. Consequently, we} are interested in fluctuations about this background governed by the charge conservation equation,
	\begin{eqnarray}
        \label{Eq:ChargeConsGeneric}
		\partial_{t} \delta \rho + \vec{\nabla} \cdot \delta \vec{J} 
 = 0 \; ,
        \end{eqnarray}
where $\delta \rho$ is a linearised fluctuation of the charge density and $\delta \vec{J}$ is a fluctuation of the corresponding spatial charge current. \new{Crucially, this charge conservation equation \eqref{Eq:ChargeConsGeneric} is also coupled to a second dynamical equation for the evolution of the spatial charge current $\delta \vec{J}$, that is assumed to be an almost conserved quantity. Unlike standard hydrodynamics, the spatial charge current $\delta \vec{J}$ is not expressed as a constitutive relation in terms of $\delta \rho$ but it is an independent thermodynamic variable. The rationale for retaining the non-conserved current in the quasi-hydrodynamic description is based on the assumption of a parametric separation of scales between the almost conserved current $\delta \vec{J}$ and the other non-conserved microscopic excitations. In particular, we assume that the current $\delta \vec{J}$ relaxes with a rate $\lambda$ that is parametrically smaller than the relaxation rate of the other microscopic modes $\Lambda$, \textit{i.e.} $\lambda \ll \Lambda$ (see Fig.1 in \cite{Baggioli:2023tlc} and related discussion). We notice that without this assumption, the theory would either reduce to only Eq.~\eqref{Eq:ChargeConsGeneric} or become intractable due to the proliferation of an infinite number of microscopic modes. As we will see later, the microscopic probe brane model will fit in this regime only in the limit of large charge density.} 

\new{It is also important to note here that the existence of thermodynamics in systems with flowing current is non-trivial. We will not concern ourselves overly with this issue here as we have a UV-complete model that displays the necessary behaviour. Moreover, we will not make use of an entropy current (or its relatives) to constrain transport coefficients in our effective quasihydrodynamics.}

\new{Given this, in the spirit of effective field theory, we write down all possible terms in a derivative expansion that respect the relevant symmetries of the system (namely spatial rotation invariance about the background electric field vector). Up to but not including order two in derivatives and fluctations, the most generic expression for the relaxation of a spatial charge current one can write down takes the form}
\begin{widetext}
    \begin{align}
    \label{Eq:GenericLinearised}
          & \partial_{t} \delta J^{i} + \partial_{j} \left( v_{\parallel}^{2} \frac{E^{i} E^{j}}{\vec{E}^2} \delta \rho + v_{\perp}^{2} \Pi^{ij} \delta \rho + \left( \eta_{1} + \eta_{2} \right) \frac{E^{i} E^{j}  E_{k}}{\vec{E}^{2}} \delta J^{k}  + \left( \eta_{2} +\eta_{3} \right) E^{(i} \Pi\indices{_{k}^{j)}} \delta J^{k}  \right.  \nonumber \\
          & \left. \hphantom{\partial_{t} \delta J^{i} + \partial_{j} \left( \right.} + \left( \eta_{2} - \eta_{3} \right) E^{[i} \Pi\indices{_{k}^{j]}} \delta J^{k} + \eta_{4} \Pi\indices{^{ij}} \delta J^{k} E_{k} \right) - \alpha E^{i} \delta \rho + \frac{1}{\tau_{\parallel}} \frac{E^{i} E^{j}}{\vec{E}^2} \delta J_{j} +\frac{1}{\tau_{\perp}} \Pi^{ij} \delta J_{j} 
            \nonumber \\
	     &  - \chi_{\parallel} \frac{E^{i} E^{j}}{\vec{E}^2} \delta E_{j} - \chi_{\perp} \Pi^{ij} \delta E_{j} + \chi_{B} \Pi\indices{^{i}_{j}} \epsilon^{jkl} E_{k} \delta B_{l}  + \zeta_{\parallel} \frac{E^{i} E^{j}}{\vec{E}^2} \partial_{t} \delta E_{j} + \zeta_{\perp} \Pi^{ij} \partial_{t} \delta E_{j} \nonumber \\
   & + \theta_{\parallel} \frac{E^{i} E^{j}}{\vec{E}^2} E^{k} \partial_{k} \delta E_{j} + \theta_{\perp} \Pi^{ij} E^{k} \partial_{k} \delta E_{j} + \theta_{\otimes} E^{i} \Pi^{jk} \partial_{j} \delta E_{k} + \theta_{\bullet} \Pi^{ij} E^{k} \partial_{j} \delta E_{k} \nonumber \\
   & + \frac{\theta_{B}}{\vec{E}^2} \Pi\indices{^{i}_{j}} \epsilon^{jkl} E_{k} E^{m} \partial_{m} \delta B_{l} + \theta_{b} E^{i} \epsilon^{jkl} E_{k} \Pi\indices{_{l}^{m}} \partial_{m} \delta B_{j}
        + \theta_{\beta} \epsilon^{ijk} E_{j} \Pi\indices{_{k}^{l}} E^{k} \partial_{l} \delta B_{k} = 0 + \mathcal{O}(\partial^2,\delta^2) \; , 
   \end{align}
\end{widetext}
\new{where all the unknown transport coefficients are functions of chemical potential and the modulus of the electric field. Here, $\mathcal{O}(\partial^2,\delta^2)$ stands for the neglected corrections which are higher order in derivatives ($\partial$) or fluctuations $(\delta)$. We notice that this equation is all orders in the electric field $\vec{E}$ since all the transport coefficients are arbitrary functions of its amplitude. Moreover,}
    \begin{eqnarray}
        \delta B_{i} = \epsilon\indices{_{i}^{jk}} \partial_{j} \delta a_{k} \;  ,
    \end{eqnarray}
 \new{with $a_k$ a background gauge field and $\Pi^{ij}= \delta^{ij} - \frac{E^i E^j}{\vec{E}^2}$.  in our derivative counting, $\delta \vec{J}$ is order zero in derivatives (similarly to $\vec{J}$). Also $E^{(i} \Pi\indices{_{k}^{j)}}$ and $E^{[i} \Pi\indices{_{k}^{j]}}$ indicate respectively the symmetric and antisymmetric parts of the tensor $E^{i} \Pi\indices{_{k}^{j}}$ with respect to the indices $i,j$.}
 
\new{Imposing the Onsager relations (that are a posteriori validated numerically) upon this expression constrains many of the background field terms but does not constrain the transport coefficients \cite{Amoretti:2023vhe}, leaving the following quantities undetermined:
    \begin{align}
    \begin{split}
        &v_{\parallel}^2 \; , \; \; v_{\perp}^2 \; , \; \; \alpha \; , \; \; \tau_{\parallel} \; , \; \; \tau_{\perp} \; , \; \; \chi_{\parallel} \; , \; \; \chi_{\perp} \; , \\
        &\;\theta_{\parallel}  \; , \; \; \theta_{\beta}  \; , \; \; \theta_{b} \; , 
       \; \eta_{1} \; , \; \; \eta_{2} \; , \; \; \eta_{3} \; , \; \; \eta_{4} \; .
    \end{split}
    \end{align}
Consistency in the $\vec{E} \rightarrow\vec{0}$ limit \cite{Chen:2017dsy} requires that
    \begin{subequations}
    \label{Eq:LowELimit}
    \begin{eqnarray}
       &\;& \left. v_{\parallel}^{2} \right|_{\vec{E} \rightarrow \vec{0}} = \left. v_{\perp}^{2} \right|_{\vec{E} \rightarrow \vec{0}} = v^2 \; , \\
       &\;&  \left. \tau_{\parallel} \right|_{\vec{E} \rightarrow \vec{0}} = \left. \tau_{\perp} \right|_{\vec{E} \rightarrow \vec{0}} = \tau \; , \\
       &\;&  \alpha, \eta_{1} , \eta_{2} , \eta_{3}, \eta_{4} \stackrel{\vec{E} \rightarrow 0}{\sim} \mathcal{O}(E^{0}) \mathrm{\; or \; higher \; powers} \; . \qquad
    \end{eqnarray}
    \end{subequations}
}
\new{ The complete equation given in Eq.~\eqref{Eq:GenericLinearised} is too complex for an initial study. Consequently, we will first take $\vec{E}$ to be small in amplitude (although not in derivatives). In other words, we will assume that the amplitude of $\vec{E}$ is small (compared to the other scales in the system such as the charge density $\rho$). By working up to order one in the amplitude of the electric field,} the resultant simplified version of Eq.~\eqref{Eq:GenericLinearised}, making use of Eq.~\eqref{Eq:LowELimit}, is:
    \begin{eqnarray}
                  \label{Eq:TransportCoeffs}
		&\;& \partial_{t} \delta J^{i} + \partial_{j} \mathcal{T}^{ij} - \alpha E^{i} \delta \rho + \frac{1}{\tau} \delta J^{i} = 0  \;,  \qquad
   \end{eqnarray}
where
\begin{align}
   \mathcal{T}^{ij}=&\,v^{2} \delta^{ij} \delta \rho + \left( \eta_{1} + \eta_{2} \right) \frac{E^{i} E^{j}  E_{k}}{\vec{E}^{2}} \delta J^{k}  + \left( \eta_{2} +\eta_{3} \right) \nonumber\\
   & E^{(i} \Pi\indices{_{k}^{j)}} \delta J^{k} + \left( \eta_{2} - \eta_{3} \right) E^{[i} \Pi\indices{_{k}^{j]}} \delta J^{k}   \vphantom{\frac{E^{i} E^{j}  E_{k}}{\vec{E}^{2}}} \nonumber \\
   & + \eta_{4} \Pi\indices{^{ij}} \delta J^{k} E_{k} \; .
\end{align}
Eq.~\eqref{Eq:TransportCoeffs} is similar to the Drude equation for electron transport. Importantly, it contains a driving term proportional to the electric field $\vec{E}$ and a dissipation term controlled by the relaxation time $\tau$.

{\ By simple manipulations, we find
    \begin{subequations}
    \begin{eqnarray}
        \label{Eq:DoubleDeriv}
        &\;& \tau \frac{\partial^2 \delta \rho}{\partial t^{2}} + \frac{\partial \delta \rho}{\partial t} + \vec{v}_\mathrm{drift} \cdot \nabla \delta \rho - D \nabla^2 \delta \rho = \mathcal{R}
    \end{eqnarray}
where
    \begin{align}
        & \vec{v}_\mathrm{drift} = - \alpha \tau \vec{E} \; , \qquad D = \tau v^2 \; , \\
        & \mathcal{R} = \tau \eta_{4} \nabla_{\perp}^2 ( \vec{E} \cdot \delta \vec{J}) + \tau \left( \eta_{1} + \eta_{2} \right) \frac{( \vec{E} \cdot \nabla)^2}{\vec{E}^{2}} \vec{E} \cdot \delta \vec{J} \nonumber \\
        & \hphantom{\mathcal{R} =} + \tau \left( \eta_{2} + \eta_{3} \right) (\vec{E} \cdot \nabla) (\nabla_{\perp} \cdot \delta \vec{J} ) \label{ttt}
    \end{align}
    \end{subequations}
and $(\nabla_{\perp})^{i} = \Pi^{ij} \nabla_{j}$. If we allow for the fact that the electric field couples directly to the charge density in the drift term, as compared to the temperature which couples through a time derivative, then Eq.~\eqref{Eq:DoubleDeriv} is formally similar to Eq.~\eqref{eq:cattaneo} with $\mathcal{R}$ given in Eq.~\eqref{ttt} \new{and such that the velocity of the material derivative is replaced by the drift velocity}. Regardless of these technical subtleties, the roles of $\vec{v}_{\mathrm{drift}}$ and $D$ for the charge density are physically the same as those for the temperature in Eq.~\eqref{eq:cattaneo}, implying a certain degree of universality in the hydrodynamic description of NESS.}

\new{In the presence of fluctuations of the background fields, $(\delta \vec{E},\delta \vec{B})$, the associated retarded charge and current correlators can be obtained up to contact terms from the following expressions:
    \begin{subequations}
    \begin{align}
        & \langle \rho \rho \rangle_{\mathrm{R}}(\omega,\vec{k}) = \lim_{\delta A \rightarrow 0} \frac{\delta \langle \rho \rangle_{\delta A}}{\delta A_{t}} \; , \\
        & \langle J^{i} \rho \rangle_{\mathrm{R}}(\omega,\vec{k}) = \lim_{\delta A \rightarrow 0} \frac{\delta \langle J^{i} \rangle_{\delta A}}{\delta A_{t}} \; , \\
        & \langle \rho J^{j} \rangle_{\mathrm{R}}(\omega,\vec{k}) = \lim_{\delta A \rightarrow 0} \frac{\delta \langle \rho \rangle_{\delta A}}{\delta A_{j}} \; , \\
        & \langle J^{i} J^{j} \rangle_{\mathrm{R}}(\omega,\vec{k}) = \lim_{\delta A \rightarrow 0} \frac{\delta \langle J^{i} \rangle_{\delta A}}{\delta A_{j}} \; , 
    \end{align}
    \end{subequations}
where $\delta \langle \rho \rangle_{\delta A}$ and $\delta \langle J^{i} \rangle_{\delta A}$ indicate the fluctuations of the one-point functions in the presence of the linearised perturbation of the gauge field. 
Consider} in particular the static solutions of Eqs.~\eqref{Eq:ChargeConsGeneric}-\eqref{Eq:TransportCoeffs}. By turning on a static electric field fluctuation, $\delta \vec{E} = i \vec{k} \delta A_{t}$, with $A_{t}$ the time component of the external gauge field $A_\mu$ \cite{Kovtun:2012rj}, one finds the retarded charge density correlator
    \begin{eqnarray}
        \label{Eq:ParallelZeroOmega}
        \langle \rho \rho \rangle_{\mathrm{R}}(0,\vec{k}) &=& \frac{ |\vec{k}| \chi}{v^2  |\vec{k}| + i \alpha |\vec{E}| \cos \varphi} \; , 
    \end{eqnarray}
where $\varphi$ is the angle between the wavevector and the electric field and we have ignored terms that are subleading in the $\vec{k}, \vec{E} \rightarrow \vec{0}$ limit.

{\ This result is peculiar from the perspective of standard hydrodynamics. For $\varphi=\pi/2$, in the limit of small wave-vector, we do obtain the standard hydrodynamic result $\chi_{\rho\rho}=\chi/v^2$ \cite{Chen:2017dsy} with $\chi_{\rho \rho}$ the charge susceptibility in the absence of the electric field. However, for other angles, the charge response is ``screened'' at small wave-vectors with the effect becoming maximal in the collinear limit $\varphi=0$. The upshot is that $\langle \rho \rho \rangle_{\mathrm{R}}$ vanishes, rather than attains the charge susceptibility as expected, in the hydrodynamic limit of $\vec{k} \rightarrow \vec{0}$ for all angles apart from $\varphi=\pi/2$.}

{\ With this stated, we can define the following length-scale $\lambda=D/|\vec{v}_{\text{drift}}|$, that is formally the inverse of the wave-vector at which the denominator of $\langle \rho \rho \rangle_{\mathrm{R}}(0,\vec{k})$ vanishes. $\lambda$ determines the scale at which the charge response $\langle \rho \rho \rangle_{\mathrm{R}}$ decays to zero and quantifies the relative weight between the diffusive and advective terms in the dynamics of the NESS.}

{\ For $\vec{k}$ parallel to $\vec{E}$, the longest-living excitations are characterized by the following dispersion relations
    \begin{subequations}
    \label{Eq:Dispersion}
    \begin{eqnarray}
        \omega_\text{gapless} &=& \alpha \,\tau \,\vec{k} \cdot \vec{E} + \mathcal{O}(\vec{k}^2,\vec{E}^2) \; , \label{lala} \\
         \label{Eq:Quasigap}
        \omega_\text{gapped} &=& - \frac{i}{\tau} + \left( \eta_{1} + \eta_{2} - \alpha \,\tau \right) \vec{k} \cdot \vec{E} + \mathcal{O}(\vec{k}^2,\vec{E}^2) \; , \qquad
    \end{eqnarray}
in the longitudinal sector, while there is a multiplicity two gapped mode with 
    \begin{eqnarray}
        \label{Eq:QuasiPerp}
        \omega_{\perp} = - \frac{i}{\tau} + \eta_{3} \,\vec{k} \cdot \vec{E} + \mathcal{O}(\vec{k}^2,\vec{E}^2) \; , 
    \end{eqnarray}
    \end{subequations}
in the transverse sector.} 

\new{We have also derived the dispersion relations of the lowest excitations at arbitrary order in $\vec{E}^2$. To avoid clutter, those formulas are presented in Appendix \ref{app1}.}

\section{Microscopic NESS model}  Within the the holographic correspondence (or gauge-gravity duality) \cite{Ammon:2015wua}, probe brane solutions \cite{Karch:2002sh} provide a class of microscopic models that realize electrically driven NESS in the regime in which the thermal bath is much larger than the system of interest and consequently dissipation is very efficient. Applying an external electric field to drive the charged particles on the probe brane, one can realize an electrically driven NESS with a stationary current owing to dissipation into the gravitational thermal bath \cite{Liu_2020}.

Holographically, the degrees of freedom of the thermal bath are described by a black brane in an asymptotically ten-dimensional anti de-Sitter (AdS) spacetime,
	\begin{equation}
            \label{Eq:ST}
		\dd s^{2} = \frac{{\ell}^{2}}{u^{2}}\left( -f(u)\dd t^{2} + \dd\vec{x}^{2} +\frac{\dd u^{2}}{f(u)} \right) + \ell^{2}\dd\Omega_{5}^{2}, 
	\end{equation}
where $\ell$ is the AdS radius, $f(u)=1-u^{4}/u_{\rm H}^{4}$ and $\dd\Omega_{5}^{2}$ is the metric on the 5-sphere. The radial coordinate $u$ ranges from the AdS boundary ($u=0$) to the black brane horizon ($u=u_{\rm H}$), and the dual field theory spacetime coordinates are taken as $(t,\vec{x}) = (t,x,y,z)$. The temperature of the thermal bath is given by $T = 1/(\pi u_{\rm H})$.
    
We insert the D7-brane as a probe of the spacetime in \eqref{Eq:ST}, meaning we ignore its backreaction \cite{Karch:2002sh}. The dynamics of the U(1) gauge field on the D7-brane is governed by the Dirac--Born--Infeld (DBI) action,
	\begin{equation}\label{dbieq}
		S_{D7} = -T_{D7} \int \dd^{8}\xi \sqrt{-\det \left( g_{ab} + 2\pi\alpha' F_{ab} \right)},
	\end{equation}
where $T_{D7}$ is the D7-brane tension, $\xi^{a}$ are the worldvolume coordinates, $g_{ab}$ is the induced metric, and $F_{ab}=\partial_{a}A_{b}-\partial_{b}A_{a}$ is the field strength of the U(1) gauge field.
	
For our purposes, we introduce a non-trivial background gauge field: $A_{t}=A_{t}(u)$ and $A_{x}=-Et+h(u)$, where $E$ corresponds to the external electric field along $x$-direction at the AdS boundary. This background gauge field describes a dual theory \cite{Karch:2007pd} with a charge density $\rho$ and current density $J$, whose explicit forms are given in Appendix \ref{app1}. In this setup, an effective horizon $u_{*}(T,E)$ emerges outside the black hole horizon as a consequence of the existence of a non-equilibrium state.
	Evaluating $\rho$ and $J$ at the effective horizon, we can derive the conductivity as a nonlinear function of $E$ in closed form,
	\begin{equation}
		\frac{\sigma_{\mathrm{DC}}}{\pi T} = \frac{\tilde{J}}{\tilde{E}} = \left(\frac{C^{2}\tilde{\rho}^{2}}{1+C^{2}\tilde{E}^{2}} + {\cal{N}}C^{4} \sqrt{1+C^{2}\tilde{E}^{2}}  \right)^{1/2},
		\label{eq:conductivity}
	\end{equation}
	where ${\cal{N}}=T_{D7}(2\pi^{2})\ell^{8} = 2\lambda' N_{c}/(2\pi)^{4}$ and $C=2\pi\alpha'/\ell^{2}$, in terms of the number of colors $N_c$, the t'Hooft coupling $\lambda'$ and the string coupling $\alpha'$.
    Each quantity is scaled by temperature: $(\tilde{J},\tilde{E},\tilde{\rho})\equiv(J/(\pi T)^{3},E/(\pi T)^{2}, \rho/(\pi T)^{3})$. In the following analysis, we set ${\cal{N}}=1$ and $C=1$ for simplicity.

 On top of this non-equilibrium steady state background, we now consider the fluctuations $A_{\mu} \to A_{\mu}+ \delta A_{\mu}(t,x,y,u)$, where we have chosen to make the fluctuations independent of the $z$-coordinate owing to the rotational symmetry in the ($y,z$)-plane. We also use a gauge in which $A_{u}=0$ so that the fluctuation of $A_{z}$ decouples.
	We then Fourier transform all the fluctuations, $\delta A_{\mu}(t,\vec{x},u) = (2\pi)^{-4}\int \dd \omega \dd^{3}k \, e^{-i\omega t+i \vec{k}\cdot \vec{x}}a_{\mu}(\omega,\vec{k},u)$ and we write $\vec{k}\cdot \vec{x} = \abs{\vec{k}}\left(x\cos\varphi + y\sin\varphi \right)$, where $\varphi$ corresponds to the angle between the momentum and background electric field (as in Eq.~\eqref{Eq:ParallelZeroOmega}).
 Finally, we rescale the frequency and momentum by temperature as $\left( \tilde{\omega}, \tilde{k} \right) = \left(\omega/(\pi T), |\vec{k}|/(\pi T) \right)$.
	
Subsequently, following the holographic dictionary \cite{Son:2002sd}, we impose in-going wave boundary conditions \cite{Mas:2009wf}, and compute the Green's functions \cite{Kaminski:2009dh} in the standard manner (more details in Appendix \ref{app2}). These models, beside having a consistent low energy NESS, are known to have some remarkable properties including an emergent form the fluctuation-dissipation theorem \cite{Sonner:2012if}. In this work however, we shall not have need to consider statistical forces, hydrodynamics on its own is enough for a steady state.

\begin{figure}
    \centering
    \includegraphics[width=0.85\linewidth]{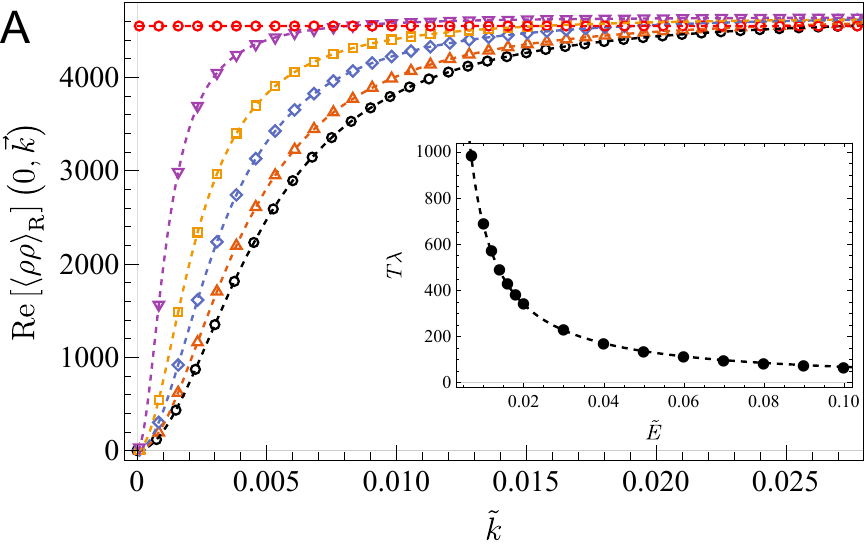}
    \includegraphics[width=0.85\linewidth]{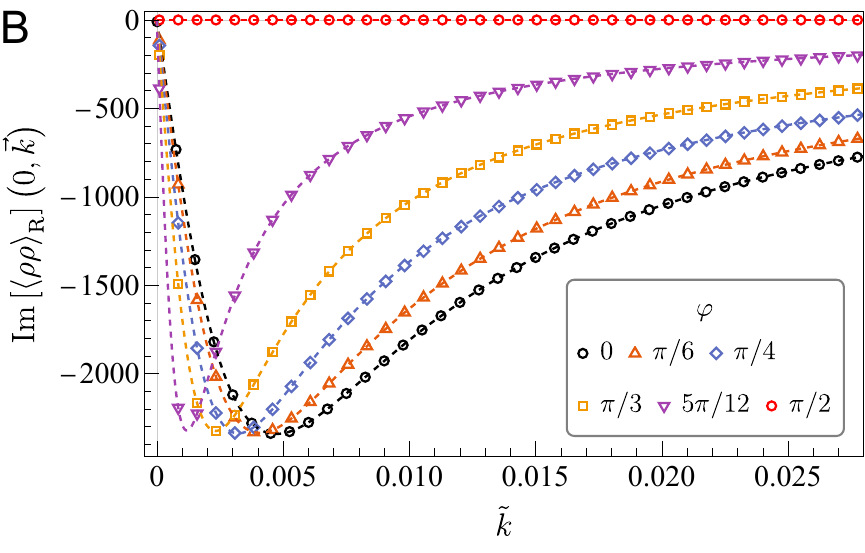} 
    \caption{Static charge response $\langle \rho \rho \rangle_{\mathrm{R}}(0,\vec{k})$ for different values of the angle $\varphi$ with $\tilde{\rho}=10^{5}$ and $\tilde{E}=0.1$. 
    \textbf{(A)} The real part and \textbf{(B)} imaginary part are shown.
    All lines are the predictions of RHT while points are holographic data from the microscopic model. The inset shows the renormalized lengthscale $\lambda$ as a function of the dimensionless electric field $\tilde E$. }
    \label{fig:1}
\end{figure}

\section{Pushing hydrodynamics out of equilibrium}
In the large charge density limit $\tilde \rho \gg 1$, in which our relaxed hydrodynamic theory applies, the values of $\tau$, $\chi$ and $v^2$ are known analytically \cite{Chen:2017dsy},
\begin{equation}\label{eq:analitic}
    \frac{1}{\tau} = \frac{(4\pi T)^{2}}{16 c \rho^{1/3}} , \qquad
    \chi_{\rho\rho} = \frac{\rho^{2/3}}{c} , \qquad
    v^{2} = \frac{1}{3},
\end{equation}
with $c = \Gamma(\frac{1}{3})\Gamma(\frac{1}{6}) / (6\sqrt{\pi})$. Thus, in Eq.~\eqref{Eq:ParallelZeroOmega} the only remaining quantity to determine is $\alpha$. This can be done in two ways: by extracting the complex-$\vec{k}$ position of the pole or, by fitting our hydrodynamic correlator to the holographic data as in Fig.~\ref{fig:1}. Both methods agree to within numerical precision.

 \begin{figure}[tbp]
    \centering
    \includegraphics[width=0.85\linewidth]{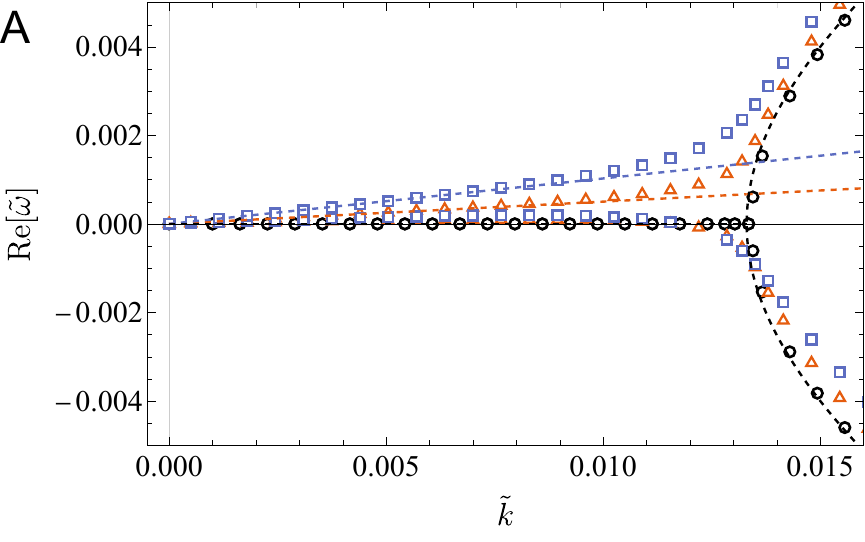}\qquad
    \includegraphics[width=0.85\linewidth]{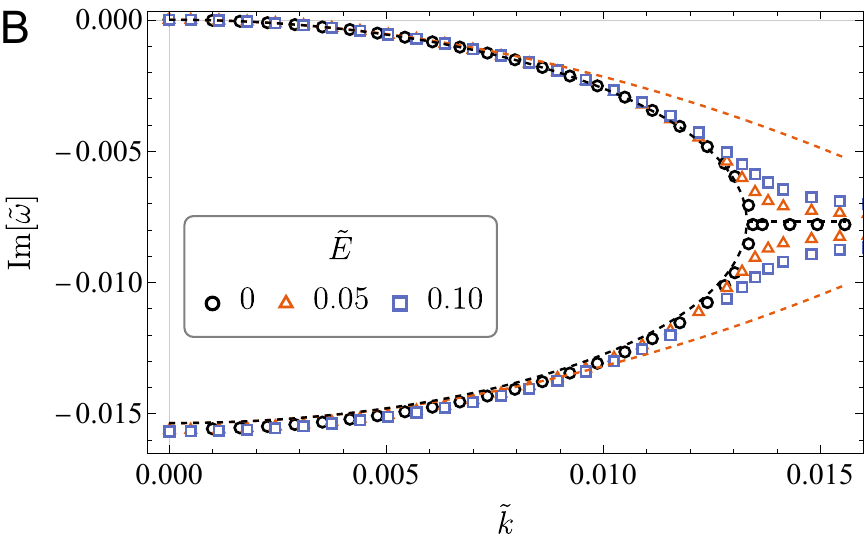}
    \caption{The dispersion $\tilde \omega(\tilde k)$ of the lowest excitations in the NESS in the collinear limit $\varphi=0$. \textbf{(A)} The real part and \textbf{(B)} imaginary part of the frequency are shown. Different colors correspond to different strength of the dimensionless electric field $\tilde E$. The symbols are the data from the microscopic model and the dashed lines are the predictions of RHT.}
    \label{fig:2}
\end{figure}

In Fig.~\ref{fig:1} we have chosen a relatively large value of the electric field, $\tilde E = 0.1$, to emphasise that our effective description applies also in the large electric field limit, on the condition that the charge density is sufficiently big. The reason for this is that, as long as $\rho$ is sufficiently large compared to $\vec{E}$, the gapped modes are very long-lived and the quasi-hydrodynamic (or relaxed approximation) holds. 
\begin{figure*}[ht]
\centering
\includegraphics[width=0.23\linewidth]{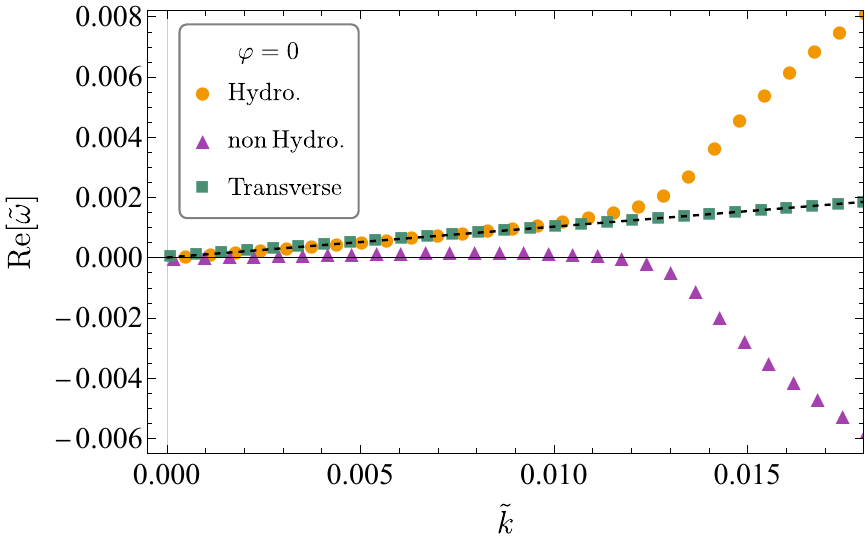}
\includegraphics[width=0.23\linewidth]{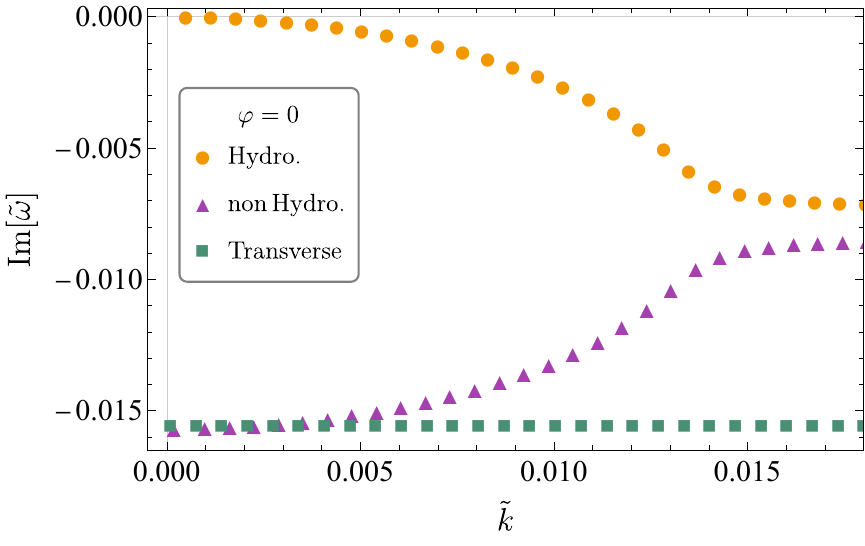}
\includegraphics[width=0.23\linewidth]{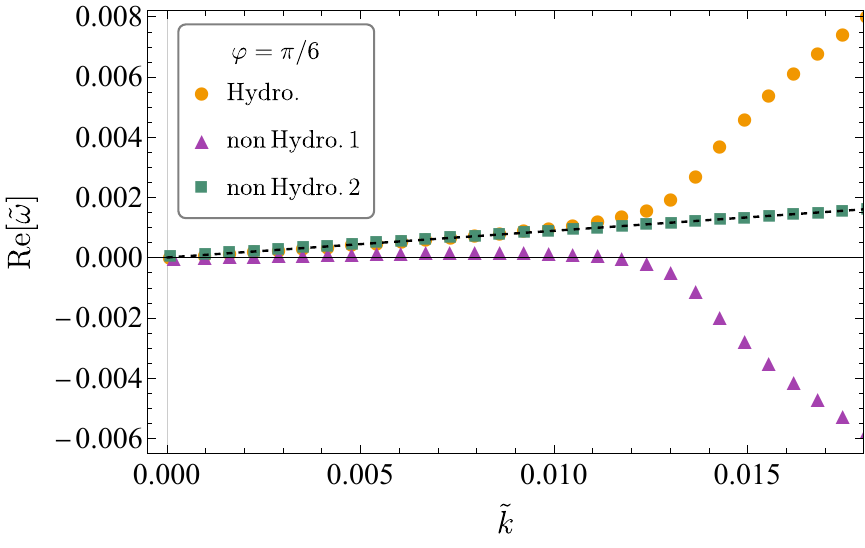}
\includegraphics[width=0.23\linewidth]{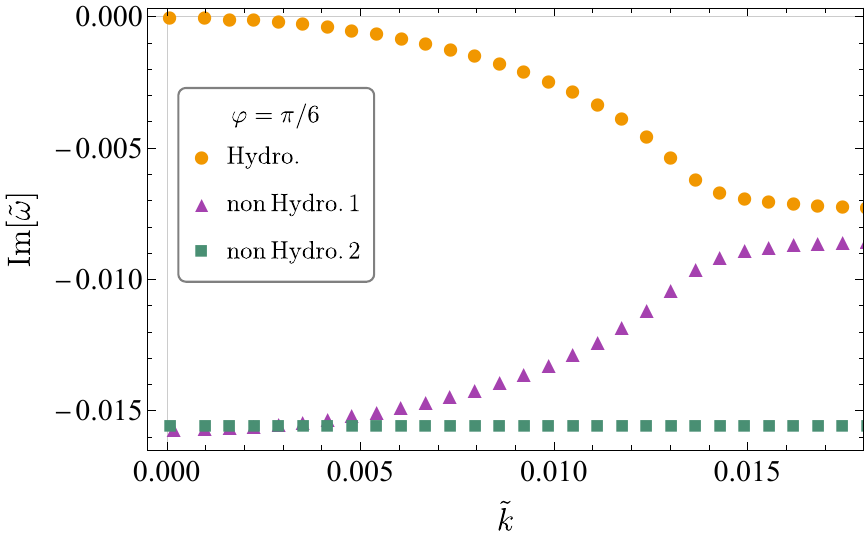}
\includegraphics[width=0.23\linewidth]{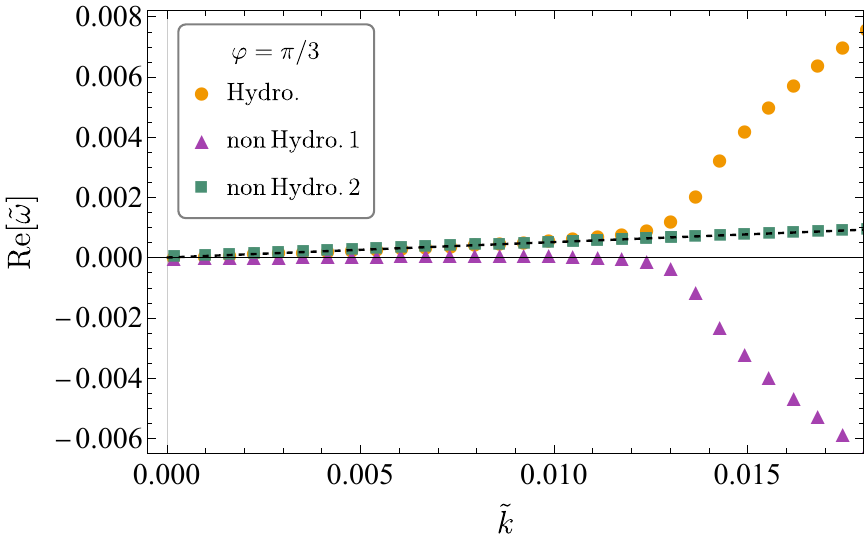}
\includegraphics[width=0.23\linewidth]{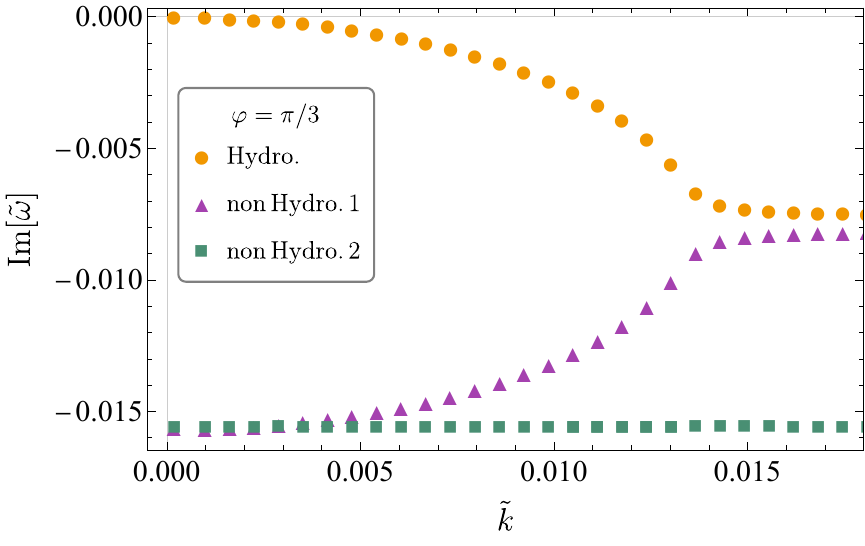}
\includegraphics[width=0.23\linewidth]{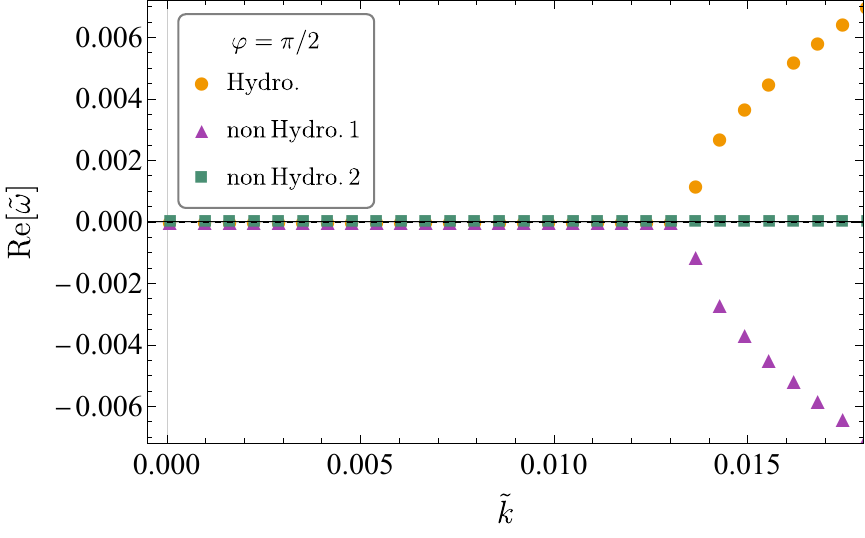}
\includegraphics[width=0.23\linewidth]{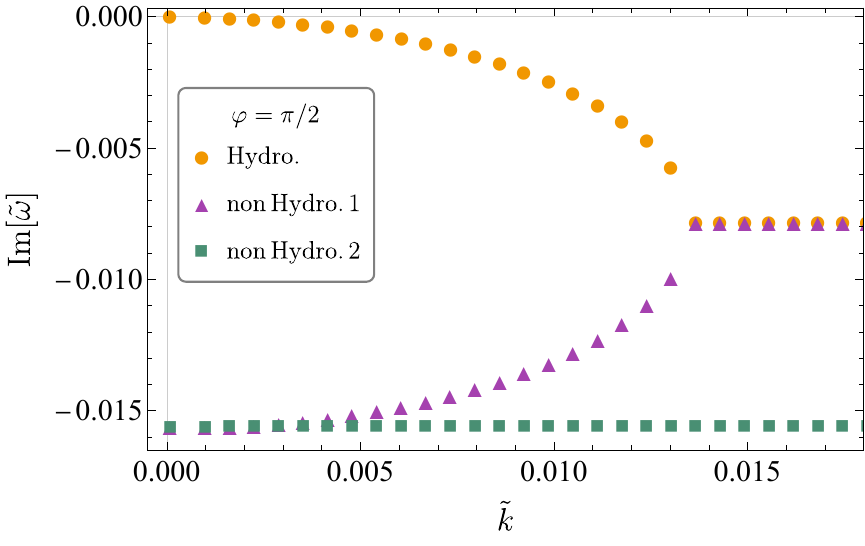}
\caption{The real and imaginary  parts of the dispersion relations for $\varphi = 0, \pi/6, \pi/3, \pi/2$ with $\tilde{E}=0.1$ and $\tilde{\rho}=10^{5}$ fixed. The black dashed lines correspond to $\mathrm{Re}\left[\tilde \omega \right]= v_{\rm drift} \tilde{k} \cos \varphi $.}
\label{fig:DispGeneral}
\end{figure*}

This suggests that the validity of the relaxed hydrodynamic theory around the NESS does not depend on the strength of the driving nor on the value of the background flow, but it is rather controlled by the relaxation time $\tau$ that corresponds to the lifetime of the first non-hydrodynamic mode, or non-conserved quantity.

We remark that the curves in Fig.~\ref{fig:1} are not independent fits. On the contrary, we use one curve at a specific angle $\varphi$ to set $\alpha$ and then all the other curves are parameter-free predictions of the RHT.

The inset of panel (A) in Fig.~\ref{fig:1} presents the numerical results for the screening length $\lambda$. The prediction of RHT is in good agreement with the numerical data and the lengthscale $\lambda$ decreases monotonically with the value of the electric field. This is a direct manifestation of the fact that increasing the electric field leads to dynamics that are more and more dominated by the drift term rather than thermal diffusive motion.

{\ We then move to consider the dynamical behavior of the system by studying the dispersion relation of the low-energy collective excitations around the NESS. The theoretical predictions are presented in Eqs.~\eqref{lala}-\eqref{Eq:Quasigap} and involve four parameters: $\tau,\alpha,\eta_1,\eta_2$. In the limit of large charge density, $\tau$ is known analytically, Eq.~\eqref{eq:analitic}, while $\alpha$ is already determined from the static charged correlator mentioned above. Therefore, only $\eta_1$ and $\eta_2$ are unknown. Unfortunately, because of the structure of the dispersion relation in Eq.~\eqref{Eq:Quasigap}, without further information, one can determine only the combination $\eta_{1+2}\equiv\eta_1+\eta_2$.}

{\ As shown in Fig.~\ref{fig:2}, the theoretical predictions fit the numerical data in the hydrodynamic limit, \textit{i.e.}, $\tilde k \ll 1$. Our fits allow us to determine the combination $\eta_{1+2}$ and yield an interesting relation, $\eta_{1+2}=4/3 \alpha \tau$ that appears robust in the limit considered. \new{To provide a more convincing test, Fig.~\ref{fig:DispGeneral} shows the real and imaginary parts of the dispersion relations for several angles with $\tilde{E}=0.1$ and $\tilde{\rho}=10^{5}$ fixed. The black dashed lines correspond to $\mathrm{Re}\left[\tilde \omega \right]= v_{\rm drift} \tilde{k}\cos \varphi $ with $v_{\rm drift} = \alpha \tau \tilde{E}$.
The agreement between the theoretical prediction and the numerical data are good independently of the value of the angle $\varphi$, proving the validity of the hydrodynamic theory.}}

{\ We have also compared the theoretical predictions for the transverse excitation, Eq.~\eqref{Eq:QuasiPerp}, and obtained good agreement as well. By following the same procedure and matching the data to the RHT formula Eq.~ \eqref{Eq:QuasiPerp},  we find that $\eta_3$, the other unknown parameter appearing in the dispersion relation of the low-energy excitations, is given by $\eta_3=\alpha \tau$ in the limit of small electric field.} 

{\ The relations above may be a consequence of some underlying thermodynamic-like identities, however because it is currently not known what the effective free energy is - nor if it exists - we cannot check this. Normally, the free energy would fix certain transport coefficients in terms of derivatives of said free energy, which is also the reason why our model contains more free parameters than would otherwise be expected. For example, we notice that the analysis of the static charge correlator and the low-energy excitations at leading order in $\tilde E$ leaves us with two undetermined parameters in the hydrodynamic equation, \eqref{Eq:TransportCoeffs}. Firstly, we are only able to determine the sum of $\eta_1+\eta_2$; second, $\eta_4$ does not appear in the dispersion of the collective excitations in the limit $\tilde E \ll 1$. This would not be the case if the thermodynamics was known.}

{\ In order to fully validate the previously derived RHT and all the undetermined ``transport coefficients'' appearing therein, we need to push the analysis to higher order in the electric field and consider the dynamics at order $\mathcal{O}\left(\tilde E^2 \right)$. In such a limit, Eq.~\eqref{Eq:TransportCoeffs} needs to be generalized to include additional terms that are subleading at order $\tilde E$, as done in the most general Eq.\eqref{Eq:GenericLinearised}. At order $\tilde E^2$, the NESS is manifestly anisotropic and all the already discussed quantities split into a parallel contribution and a transverse contribution with respect to the electric field direction. More precisely, both the relaxation time and the velocity split into $\tau_\parallel,\tau_\perp$ and $v_\parallel,v_\perp$.}

{\ We then consider the complete set of retarded correlators $\langle \rho \rho \rangle_{\mathrm{R}}(\omega,\vec{k})$, $\langle J^{i} \rho \rangle_{\mathrm{R}}(\omega,\vec{k})$, $\langle J^{i} J^{j} \rangle_{\mathrm{R}}(\omega,\vec{k})$, that can be extracted numerically from the microscopic model and derived theoretically within RHT using the background field method. We emphasize that all these quantities are complex functions and are considered up to $\mathcal{O}\left(\tilde E^2\right)$. Using the hydrodynamic formulas, we can successfully fit all the correlators and uniquely determine all the transport coefficients appearing in the hydrodynamic theory up to order $\tilde E^2$.}

\begin{figure}[b]
    \centering
    \includegraphics[width=0.85\linewidth]{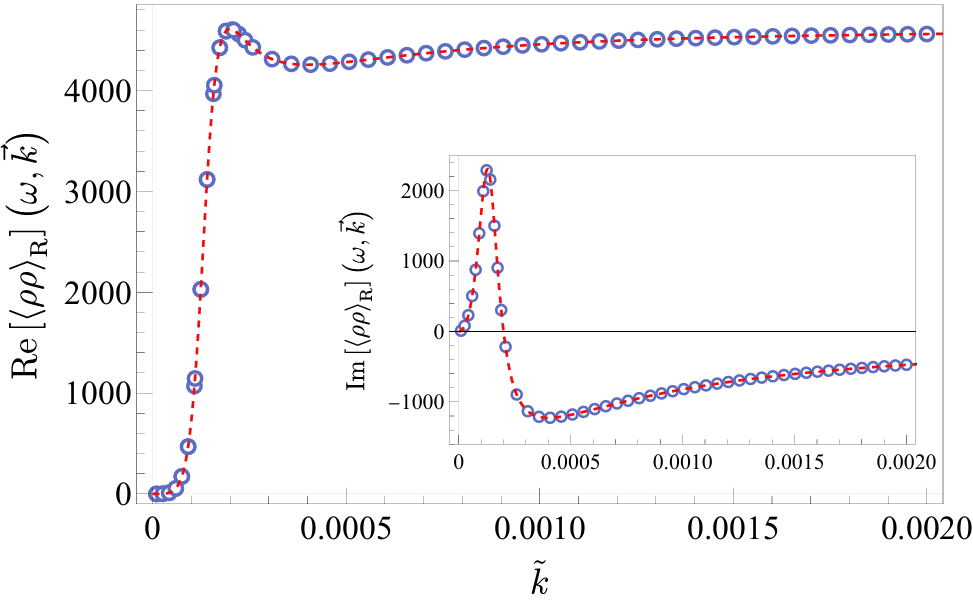}
    \caption{The charge-charge correlators at finite $\tilde{\omega}$ as a function of $\tilde{k}$ in the collinear limit $\varphi =0$ with $\tilde{\omega}=10^{-6}$, $\tilde{\rho}=10^{5}$, and $\tilde{E}=0.005$. The inset shows the imaginary part of the correlator. The points are the data from the microscopic model and the dashed lines denote the predictions of RHT.}
    \label{fig:4}
\end{figure}

{\ In Fig.~\ref{fig:4}, we provide an example of a comparison between the hydrodynamic formulas and the frequency and wave-vector dependent retarded correlator $\langle \rho \rho \rangle$. The agreement in the hydrodynamic limit is good, confirming the validity of our RHT. In Appendix \ref{app1}, we show that all the correlators are successfully reproduced by the RHT formulas.}

\section{Outlook}
We constructed a relaxed hydrodynamic theory describing the late time and long wavelength dynamics of charge and current fluctuations about electrically driven non-equilibrium charge steady states \new{valid in the limit in which the spatial current has a parametrically small relaxation rate}. Our hydrodynamic theory is accompanied by a UV complete model of these NESS, which is built using a holographic probe brane setup within the gauge-gravity duality, and utilized to corroborate our theoretical predictions. The novel hydrodynamic description necessitates the incorporation of relaxation effects -- quasi-hydrodynamics -- and applies to strongly coupled charged systems as well.

Besides the theoretical extension of hydrodynamics to NESS, our model might have an important impact in the study of charge response in strongly correlated systems or materials without quasiparticles driven by an external electric field \cite{Varnavides2023}, for which we provide clear predictions. The charge screening mechanism induced by the NESS drift velocity, and revealed by our analysis, could be in principle verified using locally resolved probe measurements (\textit{e.g.}, scanning single-electron transistors \cite{Sulpizio2019} and nitrogen vacancy center magnetometry \cite{Ku2020}). At the same time, the time-dependent dynamics could be within reach in pump-probe experiments (\textit{e.g.}, \cite{doi:10.1126/sciadv.aax3346}). It also provides the basis for searching for NESS in the presence of other quasihydrodynamic theories; for example, those with an external magnetic field \cite{Amoretti:2022acb}, with anomalous currents \cite{Amoretti:2022vxq} or exotic matter \cite{Brattan:2017yzx,Brattan:2018sgc,Brattan:2018cgk}.

Our work provides a significant step towards the applicability of hydrodynamics to non-equilibrium steady states, and it confirms once more its ``unreasonable effectiveness''. In the future, it would be interesting to expand our analysis and use the microscopic holographic model to learn more about the thermodynamics of NESS \new{presented here and its relation to those described by macroscopic fluctuation theory \cite{Bertini_2015}.}

\section*{Acknowledgments}
M.M. and M.B. acknowledge the support of the Shanghai Municipal Science and Technology Major Project (Grant No.2019SHZDZX01) and the sponsorship from the Yangyang Development Fund. A.A. and D.B. acknowledge support from the project PRIN 2022A8CJP3 by the Italian Ministry of University and Research (MUR). This project has also received funding from the European Union’s Horizon 2020 research and innovation programme under the Marie Sklodowska-Curie grant agreement No. 101030915.

\appendix

\section{Estimating the transport coefficients}\label{app1}

\begin{figure}[hb]
    \centering
    \includegraphics[width=\linewidth]{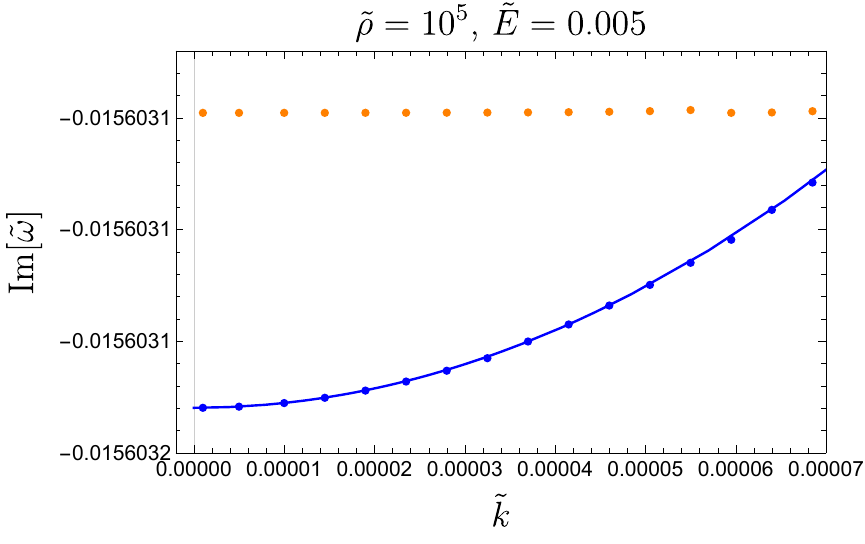}
    \caption{The dispersion relation of two non-hydro modes for small $\vec{k}$ transverse to the electric field with $\tilde{\rho}=10^{5}$ and $\tilde{E}=0.005$. The blue solid line is the prediction from the hydrodynamic theory as explained in the text.}
    \label{fig:disp3rd}
\end{figure}

{\ To obtain our transport coefficients from the holographic model, we compute the collective excitations and Green's functions coming from Eq.~\eqref{Eq:GenericLinearised} and linearised charge conservation. To avoid clutter, we will show concrete results for the following choice of parameters, $\tilde{E}=5/1000$ and $\tilde{\rho}=10^5$. Other choices of parameters lead to similar results.
We begin considering the collective excitations with $\vec{k}$ parallel to $\vec{E} = (E,0,0)$ which allow us to estimate $\tau_{\parallel}$, $\tau_{\perp}$, $\alpha$, $\eta_{1}+\eta_{2}$ and $\eta_{3}$,
    \begin{subequations}
    \begin{eqnarray}
        \omega_{\mathrm{gapless}} &=& \alpha \tau_{\parallel}  \vec{E} \cdot \vec{k} + \mathcal{O}(\vec{k}^2) \; , \\
        \omega_{\mathrm{gap}} &=& - \frac{i}{\tau_{\parallel}} + \left( \eta_{1} + \eta_{2} - \alpha \tau_{\parallel} \right) \vec{E} \cdot \vec{k} \nonumber \\
        &\;& + \mathcal{O}(\vec{k}^2) \; , \\
        \omega_{\perp} &=& - \frac{i}{\tau_{\perp}} + \eta_{3} \vec{E} \cdot \vec{k}  + \mathcal{O}(\vec{k}^2) \; , 
    \end{eqnarray}
    \end{subequations}
in the longitudinal and transverse fluctuation sectors respectively. We find
    \begin{subequations}
    \label{Eq:Transport1}
	\begin{align}
            & \alpha \tau_{\parallel} \approx \eta_{3} \approx 1.000 + \mathcal{O}(\tilde{E}^{2},\tilde{\rho}^{-\frac{1}{3}}) \; ,\\
            & \eta_{1}+\eta_{2} \approx 1.333 + \mathcal{O}(\tilde{E}^{2},\tilde{\rho}^{-\frac{1}{3}}) \; ,  \\
            & \tau_{\parallel}^{-1} \approx \left(0.713+0.358 \tilde{E}^{2}\right) \tilde{\rho}^{-1/3} \;, \\
            & \tau_{\perp}^{-1} \approx \left(0.713 +0.119 \tilde{E}^{2}\right)\tilde{\rho}^{-1/3} \; , 
	\end{align}
    \end{subequations}
    where $\tilde{E}$ and $\tilde{\rho}$ are the temperature normalised electric field magnitude and charge density respectively (defined in the main text).
Subsequently, using the hydrodynamic relation $\sigma = \tau_{\parallel}\chi_{\parallel} = \tau_{\perp} \chi_{\perp} $, we obtain
    \begin{subequations}
        \label{Eq:Transport2}
	\begin{eqnarray}
		\chi_{\parallel} &=& \sigma \tau_{\parallel}^{-1} \approx \left(0.713 +0.001\tilde{E}^{2} \right)\tilde{\rho}^{2/3} \; , \\
		\chi_{\perp} &=& \sigma \tau_{\perp}^{-1} \approx \left(0.713 -0.237\tilde{E}^{2} \right)\tilde{\rho}^{2/3} \; . 
	\end{eqnarray}
    \end{subequations}
 In the limit of small electric field, the $\mathcal{O}(\vec{E}^2)$ corrections to the displayed expressions are below the numerical precision of the leading term and will not be relevant to our initial analysis.}

{\ In addition to the above, we can obtain a relation between $\eta_{2}$ and $\eta_{4}$ from the expression for the modes when $\vec{k}$ is transverse to $\vec{E}$. In particular, choosing the wavevector perpendicular to the electric field breaks the remnant rotational invariance of the system and yields four distinct collective excitations:
\begin{widetext}
\begin{subequations}
    \begin{eqnarray}
        \omega_{\mathrm{gapless}} &=& - i \left( v_{\perp}^2 + \alpha \eta_{4} \tau_{\parallel} \vec{E}^2 \right) \vec{k}^2_{\perp} + \mathcal{O}(\vec{k}_{\perp}^4) \; , \\
        \omega_{\mathrm{gap},1} &=& - \frac{i}{\tau_{\parallel}} + \frac{i  \eta_{4} \tau_{\parallel} \tau_{\perp} (\alpha \tau_{\parallel} - \eta_{2})}{\tau_{\parallel} -\tau_{\perp}} \vec{E}^2 \vec{k}^2_{\perp} + \mathcal{O}(\vec{k}_{\perp}^4) \; , \\
        \omega_{\mathrm{gap},2} &=& - \frac{i}{\tau_{\perp}} + i \left( \left( v_{\perp}^2 + \alpha \eta_{4} \tau_{\parallel} \vec{E}^2 \right) - \frac{ \eta_{4} \tau_{\parallel} \tau_{\perp} (\alpha \tau_{\parallel} - \eta_{2})}{\tau_{\parallel} -\tau_{\perp}} \vec{E}^2 \right) \vec{k}^2_{\perp} + \mathcal{O}(\vec{k}_{\perp}^4) \; , \\
        \omega_{\mathrm{gap},3} &=& - \frac{i}{\tau_{\perp}} + \mathcal{O}(\vec{k}_{\perp}^4) \; .
    \end{eqnarray}
    \end{subequations}
    \end{widetext}
The imaginary part of one of the gapped modes, displayed in Fig.~\ref{fig:disp3rd} with orange color, is independent of $\vec{k}$ for small $\vec{k}$ in the range of $\tilde{E}$ and $\tilde{\rho}$ considered. Using this and our previously obtained values for the transport coefficients, Eqs.~\eqref{Eq:Transport1}-\eqref{Eq:Transport2}, we find the following relation
\begin{equation}
    \label{Eq:eta2relation}
    (1-\eta_{2})\eta_{4} = - 0.112 \; .
\end{equation}
Substituting this relation into the expression for the second gapless mode, displayed as the solid blue line in Fig.~\ref{fig:disp3rd}, we find consistency. We have checked this result for various large values of $\tilde{\rho}$ and small $\tilde{E}$ and obtained the same positive result.}

{\ To extract the remaining transport coefficients we compare holographic data against our Green's functions. We obtain the Green's functions through the background field method where one turns on a perturbation of the background gauge field which acts as a force in the equations of motion. In practice, for our situation, this amounts to nothing more than solving for the Fourier modes of Eq.~\eqref{Eq:GenericLinearised} plus charge conservation with the fluctuation of the gauge fields acting as a source.}

\subsection{The parameters \texorpdfstring{$v_{\parallel}^2$}{TEXT} and \texorpdfstring{$\theta_{\parallel}$}{TEXT}}

\begin{figure*}
\centering
\includegraphics[width=0.45\linewidth]{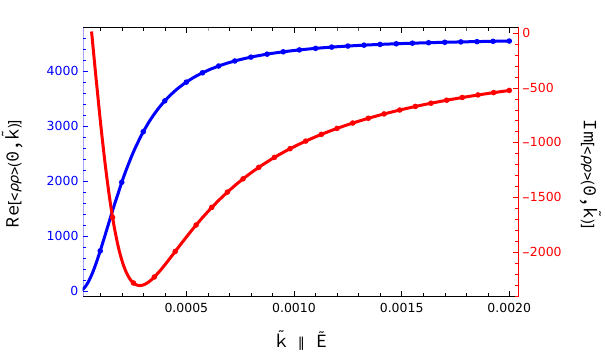} 
\includegraphics[width=0.45\linewidth]{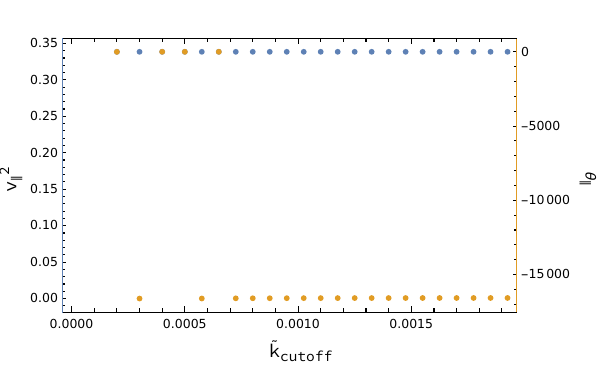} 
\includegraphics[width=0.45\linewidth]{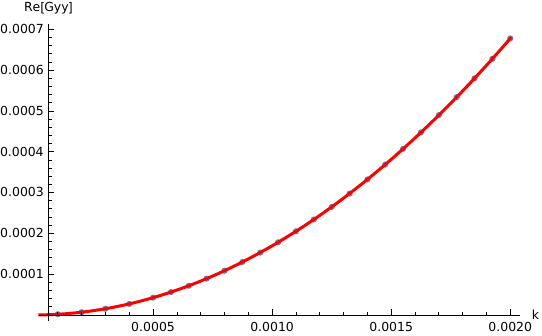} 
\includegraphics[width=0.45\linewidth]{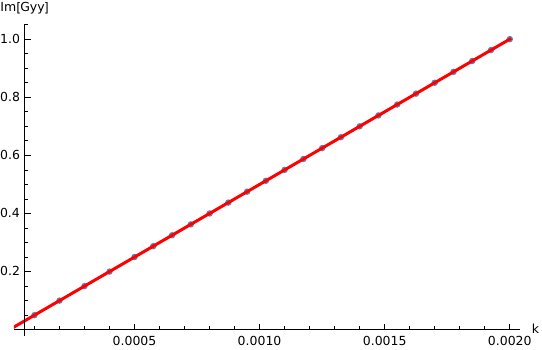} 
\includegraphics[width=0.45\linewidth]{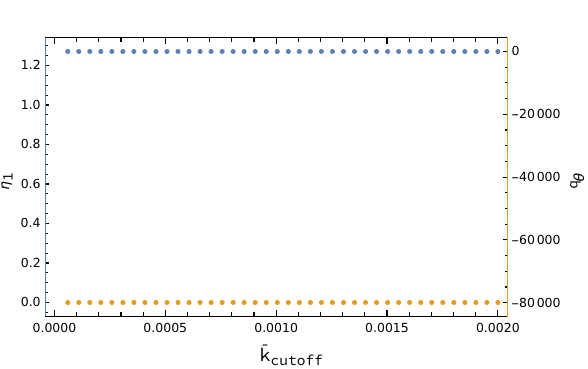} 
\caption{Various plots related to matching parameters between the holographic model and the effective quasihydrodynamic theory. 
\textbf{Top left:} a plot of the charge-charge correlator at zero frequency against wavevector for wavevector parallel to the electric field and $E=5/1000$. The solid lines are fits given by matching our quasihydrodynamic correlator, given in \eqref{Eq:Gttomega0}, against the data (represented by dots) and extracting $v_{\parallel}^2$ and $\theta_{\parallel}$ from the real part of the Green's function. 
\textbf{Middle:} Plots of the real and imaginary parts of the zero frequency current-current correlator in the $y$-direction with wavevector transverse to the electric field. The blue dots are numerical data from the holographic model and the red lines represent quadratic (real part) and linear (imaginary part) fits respectively. 
\textbf{Top right:} Our data consists of twenty points in the range $\tilde{k} \in [0, 0.002]$. We can check the stability of our fitted coefficients by reducing the largest $\tilde{k}=\tilde{k}_{\mathrm{cutoff}}$ that we include in the data-set that we fit to. We see that $v_{\parallel}^2$ is stable for all choice of $\tilde{k}$. Meanwhile, $\theta_{\parallel}$ deviates at ultra-low $\tilde{k}$ where the relevant term only makes small contributions to the correlator (c.f. Eq.~\eqref{Eq:Gttomega0}). 
\textbf{Bottom:} As in the top right panel we plot the flow of the fitted coefficient, $\eta_{1}$, against $k_{\mathrm{cutoff}}$. In this case there are forty-one points. $\theta_{b}$, also displayed, is given by an analytic expression deriving from Eq.~\eqref{Eq:Gttomega0trans}.}
\label{fig:match_parameters}
\end{figure*}

{\ At zero frequency, the retarded correlator of the charge density for $\vec{k}$ parallel to $\vec{E}$ takes the form
    \begin{eqnarray}
         \label{Eq:Gttomega0}
         \langle \rho \rho \rangle(0,\vec{k}) = \frac{|\vec{k}| (\chi_{\parallel} + |\vec{E}| |\vec{k}| \theta_{\parallel})}{v_{\parallel}^2 |\vec{k}| + i \alpha |\vec{E}|} \; .
    \end{eqnarray}
We fit the real part of this function, which turns out to be the most sensitive to the value of $\theta_{\parallel}$, to the numerical data and extract $v_{\parallel}^2$ and $\theta_{\parallel}$. The subsequent fit is displayed in the top left panel of Fig.~\ref{fig:match_parameters}. To produce the top right panel of Fig.~\ref{fig:match_parameters} we introduce a large $\vec{k}$ cut-off and then extract $v_{\parallel}^2$ and $\theta_{\parallel}$ only for data with $|\vec{k}|<k_{\mathrm{cutoff}}$. Plotting the change in the transport coefficients as the cut-off is reduced illustrates the stability in their values.}

\subsection{The parameter \texorpdfstring{$\theta_{\beta}$}{TEXT}}

{\ The $\theta_{\beta}$ transport coefficient is readily extracted from $\langle J^{z} J^{z} \rangle(0,\tilde{k})$ with $k$ perpendicular to $\vec{E}$ as
    \begin{eqnarray}
       \langle J^{z} J^{z} \rangle(0,\vec{k}) = - \theta_{\beta} \tau_{\perp} \vec{E}^2 \vec{k}^2 + \mathcal{O}(\vec{k}^3) \; .  
    \end{eqnarray}
We find (for $\tilde{E}=5/1000$) that
    \begin{eqnarray}
        \theta_{\beta} \approx 5.75 \times 10^3 \; .     
    \end{eqnarray}
}

\subsection{The parameter \texorpdfstring{$v_{\perp}^2$}{TEXT}}

{\ At the value of $\tilde{E}$ we are examining, in the stated range of $\tilde{k}$, the correlator $\langle J_{y} J_{y} \rangle$ at zero frequency with $\vec{k}$ parallel to $\vec{E}$ is approximately quadratic in the real part and linear in the imaginary part. The numerical correlator is approximately
    \begin{eqnarray}
       \langle J_{y} J_{y} \rangle(0,\vec{k}) \approx - 499.987 i \tilde{k} - 169.411 \tilde{k}^2+ \mathcal{O}(\tilde{k}^3) \; ,     
    \end{eqnarray}
which is displayed in the middle panels of Fig.~\ref{fig:match_parameters}.}

{\ In our hydrodynamic expressions, the linear piece is already fixed by the coefficients we have already determined and we find $\mathrm{Im}[\langle J_{y} J_{y} \rangle] \sim -499.994$ which represents a good agreement. The real part of the correlator can be used to fix an expression for $v_{\perp}^2$ in terms of $\eta_{1}$ and $\theta_{b}$, namely
    \begin{eqnarray}
        v_{\perp}^2 \approx 0.338 + 2.500 \times 10^{-5} \eta_{1} + 2.500 \times 10^{-10} \theta_{b} \; .
    \end{eqnarray}
Consequently, we expect the leading term to dominate $v_{\perp}^2$ for reasonable values of $\eta_{1}$ and $\theta_{b}$. This will be confirmed shortly.}

\subsection{The parameters \texorpdfstring{$\theta_{b}$}{TEXT} and \texorpdfstring{$\eta_{1}$}{TEXT}}

{\ The leading term in small $k$ for $\langle J_{t} J_{t} \rangle(0,\vec{k})$ with $\vec{k}$ perpendicular to $\vec{E}$ is
    \begin{eqnarray}
        \label{Eq:Gttomega0trans}
       \langle J_{t} J_{t} \rangle(0,\vec{k}) = \frac{\chi_{\perp}}{v_{\perp}^2 + \alpha  \vec{E}^2 \eta_{4} \tau_{\parallel}} + \mathcal{O}(\vec{E}^4 \vec{k}^2) \; . 
    \end{eqnarray}
We expect the subleading term to be strongly affected by precision errors, which bears out if we use it to fix some of our remaining transport coefficients and compare against other correlators. From \eqref{Eq:Gttomega0trans} we can determine an expression for $\theta_{b}$ in terms of $\eta_{1}$. Subsequently, we minimse the least square difference between our expression for the finite frequency correlator $\langle J_{y} J_{y} \rangle(10^{-6},\vec{k})$ with $\vec{k}$ parallel to $\vec{E}$ for $\eta_{1}$ as this seems to be the most sensitive to these values. We find
    \begin{eqnarray}
        \eta_{1} \approx 1.27 \; , \qquad \theta_{b} \approx -7.99 \times 10^4 \; . 
    \end{eqnarray}
The stability of these transport coefficients is shown in the bottom panel of Fig.~\ref{fig:match_parameters} where we plot them against varying the cut-off wavevector.}

\subsection{Summary}

\begin{figure*}
\centering
\includegraphics[width=0.45\textwidth]{figdr/GttL0.pdf} 
\includegraphics[width=0.45\textwidth]{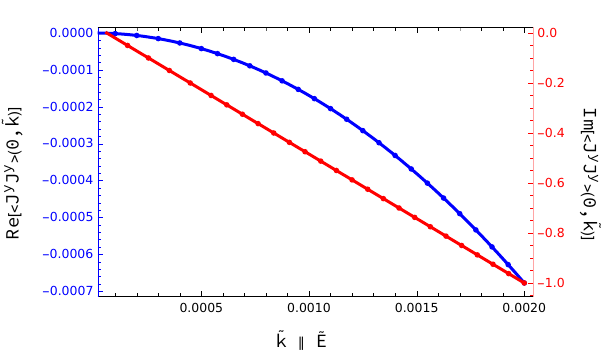} 
\includegraphics[width=0.45\textwidth]{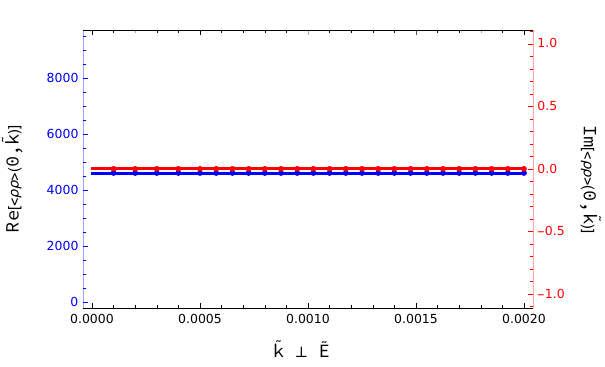} 
\includegraphics[width=0.45\textwidth]{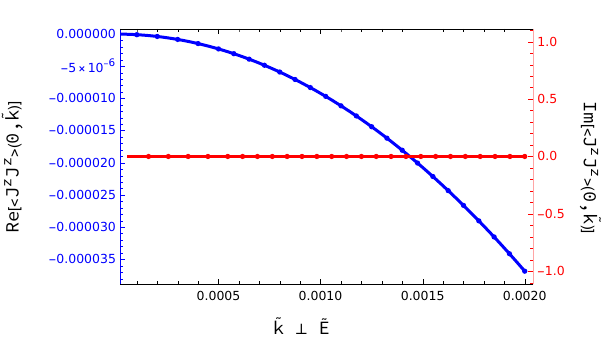}
\includegraphics[width=0.45\textwidth]{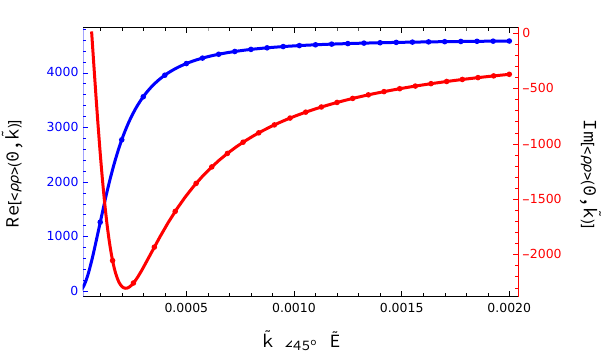} 
\includegraphics[width=0.45\textwidth]{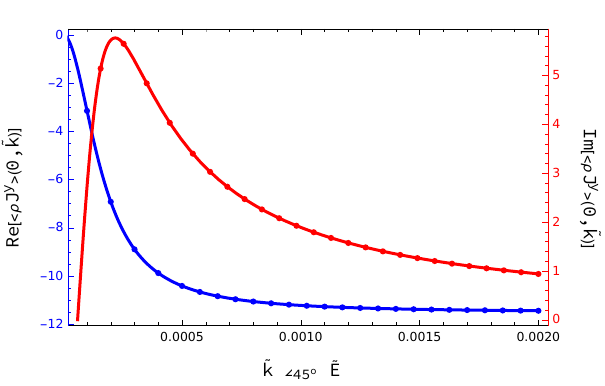}
\includegraphics[width=0.45\textwidth]{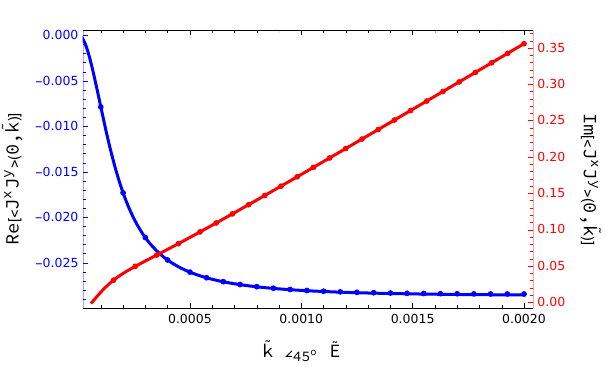} 
\includegraphics[width=0.45\textwidth]{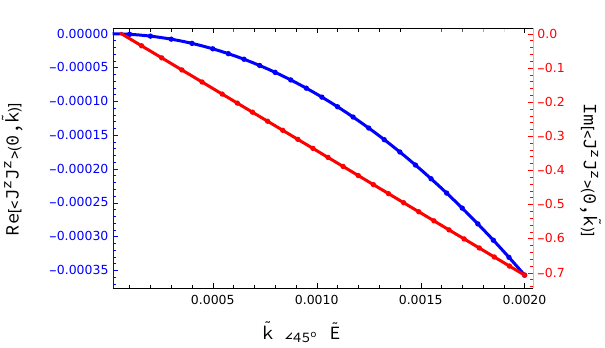} 
\caption{Non-trivial zero frequency correlators against wavevector for various angles at $\tilde{E}=5/1000$. Solid lines represent the quasihydrodynamic correlators and dots are numerical data.}
\label{fig:ZeroFrequencyCorrelators}
\end{figure*}

\begin{figure*}
\centering
\includegraphics[width=0.45\textwidth]{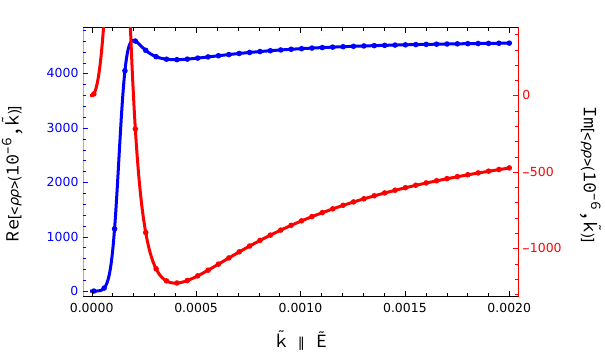} 
\includegraphics[width=0.45\textwidth]{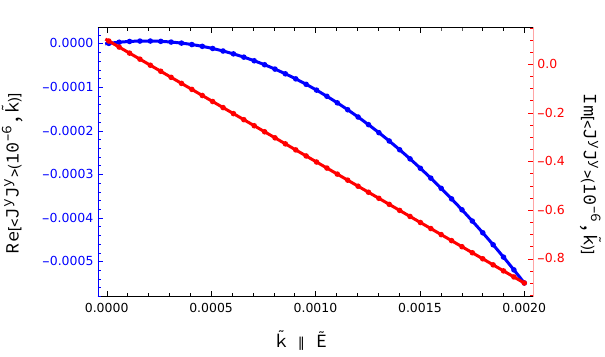} 
\includegraphics[width=0.45\textwidth]{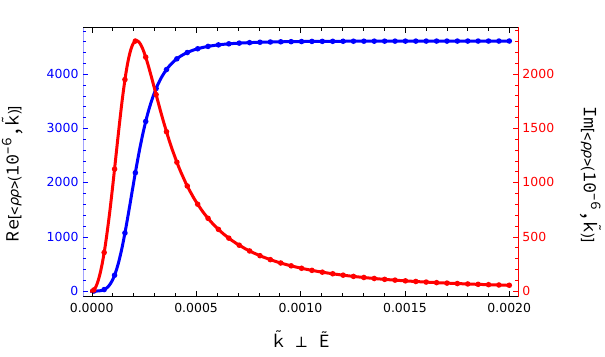} 
\includegraphics[width=0.45\textwidth]{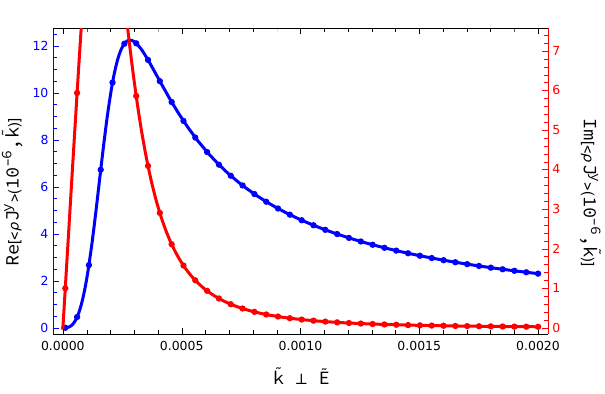} 
\includegraphics[width=0.45\textwidth]{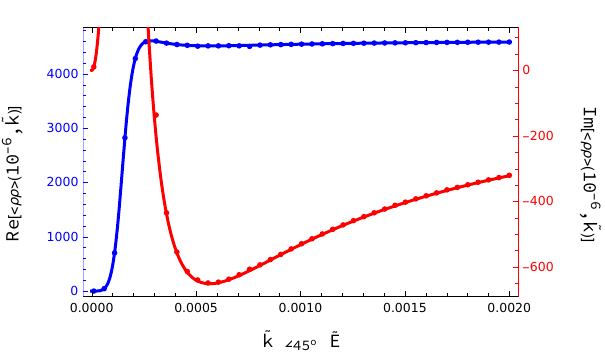} 
\includegraphics[width=0.45\textwidth]{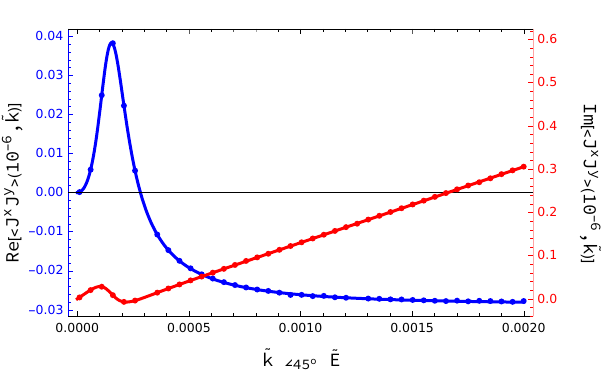} 
\includegraphics[width=0.45\textwidth]{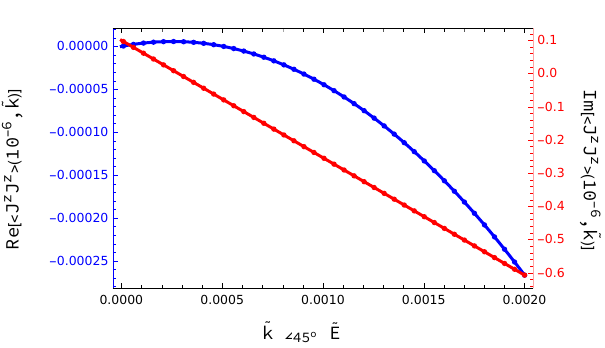} 
\caption{Correlators at $\omega = 10^{-6}$ against wavevector for various angles at $\tilde{E}=5/1000$. Solid lines represent the quasihydrodynamic correlators and dots are numerical data.}
\label{fig:NonZeroFrequencyCorrelators}
\end{figure*}

{\ The values for the transport coefficients and background field couplings with $\tilde{E} = 5/1000$ are approximately
\begin{widetext}
    \begin{align}
        & \tau_{\parallel}, \tau_{\perp} \approx 64.1 \; , \; \; \alpha \approx  0.0156 \; , \; \; v_{\parallel}^2,v_{\perp}^2 \approx 0.338 \; , \; \; \chi_{\parallel}, \chi_{\perp} \approx 1.56 \times 10^3 \; , \; \; \theta_{\parallel} \approx -1.66 \times 10^4 \; , \; \; \theta_{b} \approx -7.99 \times 10^4 \; , \; \; \nonumber\\
        & \theta_{\beta} \approx 5.75 \times 10^3 \;, \; \; \eta_{1} \approx 1.27 \; , \; \; \eta_{2} \approx 0.0485 \; , \; \; \eta_{3} \approx 1.000 \; , \; \; \eta_{4} \approx -0.117 \; . 
    \end{align}
\end{widetext}
A more detailed technical analysis is necessary to fix our transport coefficients precisely. We summarise our fits in Fig.~\ref{fig:ZeroFrequencyCorrelators} for the zero frequency correlators and Fig.~\ref{fig:NonZeroFrequencyCorrelators} for the non-zero frequency ($\omega=10^{-6}$) correlators.}

\section{Linear perturbations in the holographic probe brane model} \label{app2}
{\noindent In this section, we present more details about the microscopic holographic model.}

{\ Given the ansatz for the background gauge fields, $A_{t}=A_{t}(u)$ and $A_{x} = -E t + h(u)$, the charge density and current density are explicitly written as \cite{Karch:2007pd}
	\begin{subequations}
	\begin{align}
		\rho &= \frac{ {\cal{N}} C^{2} A_{t}'}{u \sqrt{1 -C^{2}u^{4} (A_{t}'^{2} +f^{-1}E^{2} -f h'^{2}) }}, \label{eq:rho}\\
		J & = \frac{ {\cal{N}} C^{2} f h'}{u \sqrt{1 -C^{2}u^{4} (A_{t}'^{2} +f^{-1}E^{2} -f h'^{2}) }}, \label{eq:J}
	\end{align}
	\end{subequations}
	where ${\cal{N}}=T_{D7}(2\pi^{2})\ell^{8} = 2\lambda' N_{c}/(2\pi)^{4}$ and $C=2\pi\alpha'/\ell^{2}$.
In this setup, an effective horizon $u_{*}(T,E)$ emerges outside the black hole horizon as a consequence of the existence of a non-equilibrium state.
Evaluating \eqref{eq:rho}-\eqref{eq:J} at the effective horizon, we can derive the conductivity $\sigma_{\rm DC}$ as a nonlinear function of $E$ as shown in the main text.}

In the main text, we consider the following gauge field fluctuations on the steady state background,
\begin{subequations}
\begin{align}
		&A_{t} \to A_{t}(u) + \delta A_{t}(t,x,y,u), \\
		&A_{x} \to -Et+ h(u) + \delta A_{x}(t,x,y,u), \\
		&A_{y} \to  0 + \delta A_{y}(t,x,y,u).
\end{align}		
\end{subequations}
Choosing the gauge $ A_{u}=0$, we obtain the constraint for the other fluctuations $\{ \delta A_{t}, \delta A_{x} \, \delta A_{y} \}$:
\begin{widetext}
\begin{align}
		-\frac{1}{R}\bigg[
		&-g^{tt}g^{uu}(1+g^{tt}g^{xx}E^{2}+g^{xx}g^{uu}h'^{2})(\partial_{t}\delta A_{t}')
		-g^{xx}g^{uu}(1+g^{tt}g^{xx}E^{2}+g^{tt}g^{uu}A_{t}'^{2})(\partial_{x}\delta A_{x}') \nonumber\\
		&+g^{tt}g^{xx}(g^{uu})^{2} A_{t}' h' (\partial_{x}\delta a_{t}' + \partial_{t} \delta h') 
		-g^{tt}(g^{xx})^{2} g^{uu} E h' (\partial_{x}f_{tx}) 
		- (g^{tt})^{2} g^{xx}g^{uu} E A_{t}' (\partial_{t} f_{tx})
		\bigg] \nonumber\\
		&+g^{tt}g^{xx}g^{yy}g^{uu}E h' (\partial_{y}f_{ty})
		-g^{tt}g^{xx}g^{yy}g^{uu}E A_{t}' (\partial_{y} f_{xy})
		+g^{yy} g^{uu}(1+g^{tt}g^{xx}E^{2})(\partial_{y}\delta A_{y}')=0,
\label{eq:constraint}
\end{align}
\end{widetext}
where $R = 1+ g^{tt}g^{xx}E^{2} +g^{tt}g^{uu}A_{t}'^{2} + g^{xx}g^{uu}h'^{2}$ and $f_{\mu\nu} = \partial_{\mu} \delta A_{\nu} - \partial_{\nu} \delta A_{\mu}$ with $\mu,\nu = t,x,y$.
The prime denotes the derivative with respect to $u$-coordinate.
Note that $g_{ab}^{-1}=g^{ab}$ and $g_{xx}=g_{yy}=g_{zz}$ in our setup.
Then, the quadratic action for the fluctuations can be written as
\begin{equation}
	S_{(2)} = {\cal{N}}\int \dd u \,{\cal{L}}_{(2)},
\end{equation}
where
\begin{widetext}
\begin{align}
    &\begin{aligned}
		{\cal{L}}_{(2)} = 
		\frac{1}{2}\frac{(-g)^{2}}{{\cal{L}}_{(0)}^{3}} \bigg[&
        \left(1+g^{tt}g^{uu} A_{t}'^{2}+g^{xx}g^{uu}h'^{2} \right)g^{tt}g^{xx}f_{tx}^{2}
		+\left( 1+g^{tt}g^{xx}E^{2}+g^{xx}g^{uu}h'^{2}\right)g^{tt}g^{uu}\delta A_{t}'^{2} \\
		&+\left(1+g^{tt}g^{xx}E^{2}+g^{tt}g^{uu}A_{t}'^{2} \right)g^{xx}g^{uu}\delta A_{x}'^{2} \\
		&+2(g^{tt})^{2}g^{uu}g^{xx}EA_{t}' \delta A_{t}' f_{tx} 
		+2g^{tt}(g^{xx})^{2}g^{uu} E h' \delta A_{x}' f_{tx}
		+2 g^{tt}g^{xx}(g^{uu})^{2} A_{t}' h' \delta A_{t}' \delta A_{x}' \bigg] \nonumber\\
    \end{aligned} \\[\jot]
		&\hspace{2em} +\frac{1}{2}\frac{(-g)}{{\cal{L}}_{(0)}} \bigg[ \left( 1+g^{tt}g^{xx}E^{2} \right)g^{yy}g^{uu}\delta A_{y}'^{2}
		+\left(1+g^{tt}g^{uu}A_{t}'^{2} \right)g^{xx}g^{yy}f_{xy}^{2}
		+\left(1+g^{xx}g^{uu}h'^{2} \right)g^{tt}g^{yy}f_{ty}^{2} \nonumber \\
		&\hspace{7em} -2g^{tt}g^{xx}g^{yy}g^{uu}\left(E A_{t}'\delta A_{y}'f_{xy} - E h' \delta A_{y}' f_{ty} + A_{t}' h' f_{ty}f_{xy} \right) \bigg].
\end{align}
\end{widetext}
Here we denote ${\cal{L}}_{(0)}$ as the background Lagrangian density:
\begin{equation}
	{\cal{L}}_{(0)} = - \sqrt{-g} R^{1/2},
\end{equation}
where $-g = -\det (g_{ab}) = -g_{tt}g_{xx}^{3}g_{uu}$.
We omit the volume element, $\int \dd t\dd^{3}x $, then $S_{(2)}$ precisely denotes the action density. After performing the Fourier transformation, $\delta A_{\mu} = (2\pi)^{-4}\int \dd \omega \dd^{3} \vec{k}\, e^{-i\omega t + i\vec{k}\cdot \vec{x}} a_{\mu}(\omega,\vec{k},u)$ where $\vec{k}\cdot \vec{x} = k (x \cos \varphi + y \sin \varphi)$ with $k=\abs{\vec{k}}$, we obtain a set of ordinary differential equations for the fields $( a_{t},  a_{x},  a_{y})$ from the quadratic action above.
Introducing gauge-invariant quantities, ${\cal{E}}_{\rm L} \equiv  k a_{t} \cos\varphi  + \omega  a_{x}$ and ${\cal{E}}_{\rm T} \equiv  k  a_{t} \sin\varphi  + \omega  a_{x}$, and using the constraint, we can rewrite the equations of motion for ${\cal{E}}_{\rm L}$ and ${\cal{E}}_{\rm T}$.
As their explicit forms are not illuminating, we do not write them down here.

To compute the retarded Green's function, we impose the ingoing wave boundary conditions at the horizon \cite{Son:2002sd}.
In equilibrium ($E=0$), the ingoing field at the black hole horizon corresponds to
\begin{equation}
    {\cal{E}} = \left( u - u_{\rm H} \right)^{-i \frac{\omega}{4 \pi T}} \tilde{\cal{E}},
\end{equation}
where $\tilde{\cal{E}}$ is a regular function at $u=u_{\rm H}$.
Note that in equilibrium ${\cal{E}}_{\rm L}$ and ${\cal{E}}_{\rm T}$ are identical since the system is isotropic, therefore we just wrote it as ${\cal{E}}$.
In the steady state, on the other hand, we have the effective horizon outside of the black hole horizon as mentioned in the main text.
Then, we consider the Frobenius expansion at $u=u_{*}$: 
\begin{equation}
    {\cal{E}}_{\rm L/T} = \left( u-u_{*} \right)^{i \zeta} \bar{{\cal{E}}}_{\rm L/T},
\end{equation}
where $\bar{{\cal{E}}}_{\rm L/T}$ is regular at $u=u_{*}$.
Substituting this expression into the equations of motion, we obtain the characteristic equation for $\zeta$.
In the case of $\varphi =0 $, for instance, we find two roots:
\begin{equation}
    \zeta_{0} = 0, \quad \zeta_{1} = \frac{2u_{*}^3\sqrt{f} (k A_t'+ \omega h')}{4f - u_{*} f'} ,
\end{equation}
where the functions in $\eta_{1}$ are evaluated at the effective horizon $u=u_{*}$.
Here, $\zeta = 0 $ corresponds to the ingoing wave boundary condition at the effective horizon \cite{Mas:2009wf,Ishigaki:2021vyv}.

The method for computing the retarded correlators follows Ref. \cite{Kaminski:2009dh}.
In addition to the ingoing wave boundary condition, we require to determine the normalization factors at the horizon.
Solving the equations of motion with two linearly independent normalization factors, we obtain a set of linearly independent solutions, defining the solution matrix as
\begin{eqnarray}
	M(\omega,\vec{k} ,u) \equiv 	
	\begin{bmatrix}
		{\cal{E}}_{\rm L}^{(1)} & {\cal{E}}_{\rm L}^{(2)} \\
		{\cal{E}}_{\rm T}^{(1)} & {\cal{E}}_{\rm T}^{(2)}
	\end{bmatrix},
\end{eqnarray}
where ${\cal{E}}_{L/T}^{(i)}$ $(i=1,2)$ are a set of linearly independent solutions by imposing two normalization factors at the horizon.
Using these solutions, the correlator matrix  is expressed as
\begin{widetext}
\begin{equation}
	\left< J_{\cal{E}}^{\alpha} J_{\cal{E}}^{\beta} \right>(\omega,\vec{k}) = \lim_{u \to 0}
	\frac{1}{u}\frac{1}{ \omega^{2}(\omega^{2}-k^{2})}\left(
	\begin{bmatrix}
		\omega^{2}-k^{2}\sin^{2}\varphi & k^{2}\sin\varphi\cos\varphi \\
		k^{2}\sin\varphi\cos\varphi & \omega^{2}-k^{2}\cos^{2}\varphi
	\end{bmatrix}
	M'(\omega,\vec{k},u) M(\omega,\vec{k},u)^{-1}
\right)^{\alpha\beta},
\end{equation}
\end{widetext}
where $J_{\cal{E}}^{\alpha}$ denotes the dual operator of ${\cal{E}_{\alpha}}$ with $\alpha ={\rm L}, {\rm T}$.
We then obtain the aforementioned correlators,
\begin{equation}
	\left< J^{\mu} J^{\nu} \right>(\omega,\vec{k}) = 
	\frac{\partial {\cal{E}}_{\alpha}}{\partial a_{\mu}}\frac{\partial {\cal{E}}_{\beta}}{\partial a_{\nu}}\left< J_{\cal{E}}^{\alpha} J_{\cal{E}}^{\beta} \right>(\omega,\vec{k}).
\end{equation}
The dispersion relations can be obtained by computing poles in correlators, corresponding to quasi-normal modes in the bulk. To do so, we numerically determine the frequency $\omega(k)$ such that $\lim_{u \to 0} \det M(\omega,\vec{k},u) = 0$ is satisfied with a given $k$ and background parameters.

\newpage

\newpage

\begin{thebibliography}{56}%
\makeatletter
\providecommand \@ifxundefined [1]{%
 \@ifx{#1\undefined}
}%
\providecommand \@ifnum [1]{%
 \ifnum #1\expandafter \@firstoftwo
 \else \expandafter \@secondoftwo
 \fi
}%
\providecommand \@ifx [1]{%
 \ifx #1\expandafter \@firstoftwo
 \else \expandafter \@secondoftwo
 \fi
}%
\providecommand \natexlab [1]{#1}%
\providecommand \enquote  [1]{``#1''}%
\providecommand \bibnamefont  [1]{#1}%
\providecommand \bibfnamefont [1]{#1}%
\providecommand \citenamefont [1]{#1}%
\providecommand \href@noop [0]{\@secondoftwo}%
\providecommand \href [0]{\begingroup \@sanitize@url \@href}%
\providecommand \@href[1]{\@@startlink{#1}\@@href}%
\providecommand \@@href[1]{\endgroup#1\@@endlink}%
\providecommand \@sanitize@url [0]{\catcode `\\12\catcode `\$12\catcode
  `\&12\catcode `\#12\catcode `\^12\catcode `\_12\catcode `\%12\relax}%
\providecommand \@@startlink[1]{}%
\providecommand \@@endlink[0]{}%
\providecommand \url  [0]{\begingroup\@sanitize@url \@url }%
\providecommand \@url [1]{\endgroup\@href {#1}{\urlprefix }}%
\providecommand \urlprefix  [0]{URL }%
\providecommand \Eprint [0]{\href }%
\providecommand \doibase [0]{https://doi.org/}%
\providecommand \selectlanguage [0]{\@gobble}%
\providecommand \bibinfo  [0]{\@secondoftwo}%
\providecommand \bibfield  [0]{\@secondoftwo}%
\providecommand \translation [1]{[#1]}%
\providecommand \BibitemOpen [0]{}%
\providecommand \bibitemStop [0]{}%
\providecommand \bibitemNoStop [0]{.\EOS\space}%
\providecommand \EOS [0]{\spacefactor3000\relax}%
\providecommand \BibitemShut  [1]{\csname bibitem#1\endcsname}%
\let\auto@bib@innerbib\@empty
\bibitem [{\citenamefont {Varnavides}\ \emph {et~al.}(2023)\citenamefont
  {Varnavides}, \citenamefont {Yacoby}, \citenamefont {Felser},\ and\
  \citenamefont {Narang}}]{Varnavides2023}%
  \BibitemOpen
  \bibfield  {author} {\bibinfo {author} {\bibfnamefont {G.}~\bibnamefont
  {Varnavides}}, \bibinfo {author} {\bibfnamefont {A.}~\bibnamefont {Yacoby}},
  \bibinfo {author} {\bibfnamefont {C.}~\bibnamefont {Felser}},\ and\ \bibinfo
  {author} {\bibfnamefont {P.}~\bibnamefont {Narang}},\ }\bibfield  {title}
  {\bibinfo {title} {Charge transport and hydrodynamics in materials},\ }\href
  {https://doi.org/10.1038/s41578-023-00597-3} {\bibfield  {journal} {\bibinfo
  {journal} {Nature Reviews Materials}\ }\textbf {\bibinfo {volume} {8}},\
  \bibinfo {pages} {726} (\bibinfo {year} {2023})}\BibitemShut {NoStop}%
\bibitem [{\citenamefont {Halperin}\ and\ \citenamefont
  {Hohenberg}(1969)}]{PhysRev.188.898}%
  \BibitemOpen
  \bibfield  {author} {\bibinfo {author} {\bibfnamefont {B.~I.}\ \bibnamefont
  {Halperin}}\ and\ \bibinfo {author} {\bibfnamefont {P.~C.}\ \bibnamefont
  {Hohenberg}},\ }\bibfield  {title} {\bibinfo {title} {Hydrodynamic theory of
  spin waves},\ }\href {https://doi.org/10.1103/PhysRev.188.898} {\bibfield
  {journal} {\bibinfo  {journal} {Phys. Rev.}\ }\textbf {\bibinfo {volume}
  {188}},\ \bibinfo {pages} {898} (\bibinfo {year} {1969})}\BibitemShut
  {NoStop}%
\bibitem [{\citenamefont {Fleming}\ and\ \citenamefont
  {Cohen}(1976)}]{PhysRevB.13.500}%
  \BibitemOpen
  \bibfield  {author} {\bibinfo {author} {\bibfnamefont {P.~D.}\ \bibnamefont
  {Fleming}}\ and\ \bibinfo {author} {\bibfnamefont {C.}~\bibnamefont
  {Cohen}},\ }\bibfield  {title} {\bibinfo {title} {Hydrodynamics of solids},\
  }\href {https://doi.org/10.1103/PhysRevB.13.500} {\bibfield  {journal}
  {\bibinfo  {journal} {Phys. Rev. B}\ }\textbf {\bibinfo {volume} {13}},\
  \bibinfo {pages} {500} (\bibinfo {year} {1976})}\BibitemShut {NoStop}%
\bibitem [{\citenamefont {Martin}\ \emph {et~al.}(1972)\citenamefont {Martin},
  \citenamefont {Parodi},\ and\ \citenamefont {Pershan}}]{PhysRevA.6.2401}%
  \BibitemOpen
  \bibfield  {author} {\bibinfo {author} {\bibfnamefont {P.~C.}\ \bibnamefont
  {Martin}}, \bibinfo {author} {\bibfnamefont {O.}~\bibnamefont {Parodi}},\
  and\ \bibinfo {author} {\bibfnamefont {P.~S.}\ \bibnamefont {Pershan}},\
  }\bibfield  {title} {\bibinfo {title} {Unified hydrodynamic theory for
  crystals, liquid crystals, and normal fluids},\ }\href
  {https://doi.org/10.1103/PhysRevA.6.2401} {\bibfield  {journal} {\bibinfo
  {journal} {Phys. Rev. A}\ }\textbf {\bibinfo {volume} {6}},\ \bibinfo {pages}
  {2401} (\bibinfo {year} {1972})}\BibitemShut {NoStop}%
\bibitem [{\citenamefont {Lax}(1960)}]{RevModPhys.32.25}%
  \BibitemOpen
  \bibfield  {author} {\bibinfo {author} {\bibfnamefont {M.}~\bibnamefont
  {Lax}},\ }\bibfield  {title} {\bibinfo {title} {Fluctuations from the
  nonequilibrium steady state},\ }\href
  {https://doi.org/10.1103/RevModPhys.32.25} {\bibfield  {journal} {\bibinfo
  {journal} {Rev. Mod. Phys.}\ }\textbf {\bibinfo {volume} {32}},\ \bibinfo
  {pages} {25} (\bibinfo {year} {1960})}\BibitemShut {NoStop}%
\bibitem [{\citenamefont {Tremblay}\ \emph {et~al.}(1981)\citenamefont
  {Tremblay}, \citenamefont {Arai},\ and\ \citenamefont
  {Siggia}}]{PhysRevA.23.1451}%
  \BibitemOpen
  \bibfield  {author} {\bibinfo {author} {\bibfnamefont {A.~M.~S.}\
  \bibnamefont {Tremblay}}, \bibinfo {author} {\bibfnamefont {M.}~\bibnamefont
  {Arai}},\ and\ \bibinfo {author} {\bibfnamefont {E.~D.}\ \bibnamefont
  {Siggia}},\ }\bibfield  {title} {\bibinfo {title} {Fluctuations about simple
  nonequilibrium steady states},\ }\href
  {https://doi.org/10.1103/PhysRevA.23.1451} {\bibfield  {journal} {\bibinfo
  {journal} {Phys. Rev. A}\ }\textbf {\bibinfo {volume} {23}},\ \bibinfo
  {pages} {1451} (\bibinfo {year} {1981})}\BibitemShut {NoStop}%
\bibitem [{\citenamefont {Procaccia}\ \emph
  {et~al.}(1979{\natexlab{a}})\citenamefont {Procaccia}, \citenamefont {Ronis},
  \citenamefont {Collins}, \citenamefont {Ross},\ and\ \citenamefont
  {Oppenheim}}]{PhysRevA.19.1290}%
  \BibitemOpen
  \bibfield  {author} {\bibinfo {author} {\bibfnamefont {I.}~\bibnamefont
  {Procaccia}}, \bibinfo {author} {\bibfnamefont {D.}~\bibnamefont {Ronis}},
  \bibinfo {author} {\bibfnamefont {M.~A.}\ \bibnamefont {Collins}}, \bibinfo
  {author} {\bibfnamefont {J.}~\bibnamefont {Ross}},\ and\ \bibinfo {author}
  {\bibfnamefont {I.}~\bibnamefont {Oppenheim}},\ }\bibfield  {title} {\bibinfo
  {title} {Statistical mechanics of stationary states. i. formal theory},\
  }\href {https://doi.org/10.1103/PhysRevA.19.1290} {\bibfield  {journal}
  {\bibinfo  {journal} {Phys. Rev. A}\ }\textbf {\bibinfo {volume} {19}},\
  \bibinfo {pages} {1290} (\bibinfo {year} {1979}{\natexlab{a}})}\BibitemShut
  {NoStop}%
\bibitem [{\citenamefont {Procaccia}\ \emph
  {et~al.}(1979{\natexlab{b}})\citenamefont {Procaccia}, \citenamefont
  {Ronis},\ and\ \citenamefont {Oppenheim}}]{PhysRevLett.42.287}%
  \BibitemOpen
  \bibfield  {author} {\bibinfo {author} {\bibfnamefont {I.}~\bibnamefont
  {Procaccia}}, \bibinfo {author} {\bibfnamefont {D.}~\bibnamefont {Ronis}},\
  and\ \bibinfo {author} {\bibfnamefont {I.}~\bibnamefont {Oppenheim}},\
  }\bibfield  {title} {\bibinfo {title} {Light scattering from nonequilibrium
  stationary states: The implication of broken time-reversal symmetry},\ }\href
  {https://doi.org/10.1103/PhysRevLett.42.287} {\bibfield  {journal} {\bibinfo
  {journal} {Phys. Rev. Lett.}\ }\textbf {\bibinfo {volume} {42}},\ \bibinfo
  {pages} {287} (\bibinfo {year} {1979}{\natexlab{b}})}\BibitemShut {NoStop}%
\bibitem [{\citenamefont {Law}\ and\ \citenamefont {Sengers}(1989)}]{Law1989}%
  \BibitemOpen
  \bibfield  {author} {\bibinfo {author} {\bibfnamefont {B.~M.}\ \bibnamefont
  {Law}}\ and\ \bibinfo {author} {\bibfnamefont {J.~V.}\ \bibnamefont
  {Sengers}},\ }\bibfield  {title} {\bibinfo {title} {Fluctuations in fluids
  out of thermal equilibrium},\ }\href {https://doi.org/10.1007/BF01022821}
  {\bibfield  {journal} {\bibinfo  {journal} {Journal of Statistical Physics}\
  }\textbf {\bibinfo {volume} {57}},\ \bibinfo {pages} {531} (\bibinfo {year}
  {1989})}\BibitemShut {NoStop}%
\bibitem [{\citenamefont {Derrida}(2007)}]{derrida2007non}%
  \BibitemOpen
  \bibfield  {author} {\bibinfo {author} {\bibfnamefont {B.}~\bibnamefont
  {Derrida}},\ }\bibfield  {title} {\bibinfo {title} {Non-equilibrium steady
  states: fluctuations and large deviations of the density and of the
  current},\ }\href@noop {} {\bibfield  {journal} {\bibinfo  {journal} {Journal
  of Statistical Mechanics: Theory and Experiment}\ }\textbf {\bibinfo {volume}
  {2007}},\ \bibinfo {pages} {P07023} (\bibinfo {year} {2007})}\BibitemShut
  {NoStop}%
\bibitem [{\citenamefont {Bertini}\ \emph {et~al.}(2001)\citenamefont
  {Bertini}, \citenamefont {De~Sole}, \citenamefont {Gabrielli}, \citenamefont
  {Jona-Lasinio},\ and\ \citenamefont {Landim}}]{PhysRevLett.87.040601}%
  \BibitemOpen
  \bibfield  {author} {\bibinfo {author} {\bibfnamefont {L.}~\bibnamefont
  {Bertini}}, \bibinfo {author} {\bibfnamefont {A.}~\bibnamefont {De~Sole}},
  \bibinfo {author} {\bibfnamefont {D.}~\bibnamefont {Gabrielli}}, \bibinfo
  {author} {\bibfnamefont {G.}~\bibnamefont {Jona-Lasinio}},\ and\ \bibinfo
  {author} {\bibfnamefont {C.}~\bibnamefont {Landim}},\ }\bibfield  {title}
  {\bibinfo {title} {Fluctuations in stationary nonequilibrium states of
  irreversible processes},\ }\href
  {https://doi.org/10.1103/PhysRevLett.87.040601} {\bibfield  {journal}
  {\bibinfo  {journal} {Phys. Rev. Lett.}\ }\textbf {\bibinfo {volume} {87}},\
  \bibinfo {pages} {040601} (\bibinfo {year} {2001})}\BibitemShut {NoStop}%
\bibitem [{\citenamefont {Bertini}\ \emph {et~al.}(2002)\citenamefont
  {Bertini}, \citenamefont {De~Sole}, \citenamefont {Gabrielli}, \citenamefont
  {Jona-Lasinio},\ and\ \citenamefont {Landim}}]{Bertini2002}%
  \BibitemOpen
  \bibfield  {author} {\bibinfo {author} {\bibfnamefont {L.}~\bibnamefont
  {Bertini}}, \bibinfo {author} {\bibfnamefont {A.}~\bibnamefont {De~Sole}},
  \bibinfo {author} {\bibfnamefont {D.}~\bibnamefont {Gabrielli}}, \bibinfo
  {author} {\bibfnamefont {G.}~\bibnamefont {Jona-Lasinio}},\ and\ \bibinfo
  {author} {\bibfnamefont {C.}~\bibnamefont {Landim}},\ }\bibfield  {title}
  {\bibinfo {title} {Macroscopic fluctuation theory for stationary
  non-equilibrium states},\ }\href {https://doi.org/10.1023/A:1014525911391}
  {\bibfield  {journal} {\bibinfo  {journal} {Journal of Statistical Physics}\
  }\textbf {\bibinfo {volume} {107}},\ \bibinfo {pages} {635} (\bibinfo {year}
  {2002})}\BibitemShut {NoStop}%
\bibitem [{\citenamefont {Bertini}\ \emph {et~al.}(2015)\citenamefont
  {Bertini}, \citenamefont {De~Sole}, \citenamefont {Gabrielli}, \citenamefont
  {Jona-Lasinio},\ and\ \citenamefont {Landim}}]{Bertini_2015}%
  \BibitemOpen
  \bibfield  {author} {\bibinfo {author} {\bibfnamefont {L.}~\bibnamefont
  {Bertini}}, \bibinfo {author} {\bibfnamefont {A.}~\bibnamefont {De~Sole}},
  \bibinfo {author} {\bibfnamefont {D.}~\bibnamefont {Gabrielli}}, \bibinfo
  {author} {\bibfnamefont {G.}~\bibnamefont {Jona-Lasinio}},\ and\ \bibinfo
  {author} {\bibfnamefont {C.}~\bibnamefont {Landim}},\ }\bibfield  {title}
  {\bibinfo {title} {Macroscopic fluctuation theory},\ }\href
  {https://doi.org/10.1103/revmodphys.87.593} {\bibfield  {journal} {\bibinfo
  {journal} {Reviews of Modern Physics}\ }\textbf {\bibinfo {volume} {87}},\
  \bibinfo {pages} {593–636} (\bibinfo {year} {2015})}\BibitemShut {NoStop}%
\bibitem [{\citenamefont {Maxwell}(1867)}]{maxwell1867iv}%
  \BibitemOpen
  \bibfield  {author} {\bibinfo {author} {\bibfnamefont {J.~C.}\ \bibnamefont
  {Maxwell}},\ }\bibfield  {title} {\bibinfo {title} {Iv. on the dynamical
  theory of gases},\ }\href@noop {} {\bibfield  {journal} {\bibinfo  {journal}
  {Philosophical transactions of the Royal Society of London}\ ,\ \bibinfo
  {pages} {49}} (\bibinfo {year} {1867})}\BibitemShut {NoStop}%
\bibitem [{\citenamefont {Cattaneo}(1958)}]{cattaneo1958forme}%
  \BibitemOpen
  \bibfield  {author} {\bibinfo {author} {\bibfnamefont {C.}~\bibnamefont
  {Cattaneo}},\ }\bibfield  {title} {\bibinfo {title} {Sur une forme de
  l'equation de la chaleur eliminant la paradoxe d'une propagation
  instantantee},\ }\href@noop {} {\bibfield  {journal} {\bibinfo  {journal}
  {Compt. Rendu}\ }\textbf {\bibinfo {volume} {247}},\ \bibinfo {pages} {431}
  (\bibinfo {year} {1958})}\BibitemShut {NoStop}%
\bibitem [{\citenamefont {Joseph}\ and\ \citenamefont
  {Preziosi}(1989)}]{RevModPhys.61.41}%
  \BibitemOpen
  \bibfield  {author} {\bibinfo {author} {\bibfnamefont {D.~D.}\ \bibnamefont
  {Joseph}}\ and\ \bibinfo {author} {\bibfnamefont {L.}~\bibnamefont
  {Preziosi}},\ }\bibfield  {title} {\bibinfo {title} {Heat waves},\ }\href
  {https://doi.org/10.1103/RevModPhys.61.41} {\bibfield  {journal} {\bibinfo
  {journal} {Rev. Mod. Phys.}\ }\textbf {\bibinfo {volume} {61}},\ \bibinfo
  {pages} {41} (\bibinfo {year} {1989})}\BibitemShut {NoStop}%
\bibitem [{\citenamefont {Grozdanov}\ \emph {et~al.}(2019)\citenamefont
  {Grozdanov}, \citenamefont {Lucas},\ and\ \citenamefont
  {Poovuttikul}}]{Grozdanov:2018fic}%
  \BibitemOpen
  \bibfield  {author} {\bibinfo {author} {\bibfnamefont {S.}~\bibnamefont
  {Grozdanov}}, \bibinfo {author} {\bibfnamefont {A.}~\bibnamefont {Lucas}},\
  and\ \bibinfo {author} {\bibfnamefont {N.}~\bibnamefont {Poovuttikul}},\
  }\bibfield  {title} {\bibinfo {title} {{Holography and hydrodynamics with
  weakly broken symmetries}},\ }\href
  {https://doi.org/10.1103/PhysRevD.99.086012} {\bibfield  {journal} {\bibinfo
  {journal} {Phys. Rev. D}\ }\textbf {\bibinfo {volume} {99}},\ \bibinfo
  {pages} {086012} (\bibinfo {year} {2019})},\ \Eprint
  {https://arxiv.org/abs/1810.10016} {arXiv:1810.10016 [hep-th]} \BibitemShut
  {NoStop}%
\bibitem [{\citenamefont {Jain}\ and\ \citenamefont
  {Kovtun}(2024)}]{Jain:2023obu}%
  \BibitemOpen
  \bibfield  {author} {\bibinfo {author} {\bibfnamefont {A.}~\bibnamefont
  {Jain}}\ and\ \bibinfo {author} {\bibfnamefont {P.}~\bibnamefont {Kovtun}},\
  }\bibfield  {title} {\bibinfo {title} {{Schwinger-Keldysh effective field
  theory for stable and causal relativistic hydrodynamics}},\ }\href
  {https://doi.org/10.1007/JHEP01(2024)162} {\bibfield  {journal} {\bibinfo
  {journal} {JHEP}\ }\textbf {\bibinfo {volume} {01}},\ \bibinfo {pages}
  {162}},\ \Eprint {https://arxiv.org/abs/2309.00511} {arXiv:2309.00511
  [hep-th]} \BibitemShut {NoStop}%
\bibitem [{\citenamefont
  {Chandrasekharaiah}(1998)}]{chandrasekharaiah1998hyperbolic}%
  \BibitemOpen
  \bibfield  {author} {\bibinfo {author} {\bibfnamefont {D.~S.}\ \bibnamefont
  {Chandrasekharaiah}},\ }\bibfield  {title} {\bibinfo {title} {{Hyperbolic
  Thermoelasticity: A Review of Recent Literature}},\ }\href
  {https://doi.org/10.1115/1.3098984} {\bibfield  {journal} {\bibinfo
  {journal} {Applied Mechanics Reviews}\ }\textbf {\bibinfo {volume} {51}},\
  \bibinfo {pages} {705} (\bibinfo {year} {1998})}\BibitemShut {NoStop}%
\bibitem [{\citenamefont {Christov}\ and\ \citenamefont
  {Jordan}(2005)}]{PhysRevLett.94.154301}%
  \BibitemOpen
  \bibfield  {author} {\bibinfo {author} {\bibfnamefont {C.~I.}\ \bibnamefont
  {Christov}}\ and\ \bibinfo {author} {\bibfnamefont {P.~M.}\ \bibnamefont
  {Jordan}},\ }\bibfield  {title} {\bibinfo {title} {Heat conduction paradox
  involving second-sound propagation in moving media},\ }\href
  {https://doi.org/10.1103/PhysRevLett.94.154301} {\bibfield  {journal}
  {\bibinfo  {journal} {Phys. Rev. Lett.}\ }\textbf {\bibinfo {volume} {94}},\
  \bibinfo {pages} {154301} (\bibinfo {year} {2005})}\BibitemShut {NoStop}%
\bibitem [{\citenamefont {Christov}(2009)}]{CHRISTOV2009481}%
  \BibitemOpen
  \bibfield  {author} {\bibinfo {author} {\bibfnamefont {C.}~\bibnamefont
  {Christov}},\ }\bibfield  {title} {\bibinfo {title} {On frame indifferent
  formulation of the maxwell–cattaneo model of finite-speed heat
  conduction},\ }\href@noop {} {\bibfield  {journal} {\bibinfo  {journal}
  {Mechanics Research Communications}\ }\textbf {\bibinfo {volume} {36}},\
  \bibinfo {pages} {481} (\bibinfo {year} {2009})}\BibitemShut {NoStop}%
\bibitem [{\citenamefont {Bissell}(2015)}]{doi:10.1098/rspa.2014.0845}%
  \BibitemOpen
  \bibfield  {author} {\bibinfo {author} {\bibfnamefont {J.~J.}\ \bibnamefont
  {Bissell}},\ }\bibfield  {title} {\bibinfo {title} {On oscillatory convection
  with the cattaneo–christov hyperbolic heat-flow model},\ }\href
  {https://doi.org/10.1098/rspa.2014.0845} {\bibfield  {journal} {\bibinfo
  {journal} {Proceedings of the Royal Society A: Mathematical, Physical and
  Engineering Sciences}\ }\textbf {\bibinfo {volume} {471}},\ \bibinfo {pages}
  {20140845} (\bibinfo {year} {2015})}\BibitemShut {NoStop}%
\bibitem [{\citenamefont {Straughan}(2010)}]{STRAUGHAN201095}%
  \BibitemOpen
  \bibfield  {author} {\bibinfo {author} {\bibfnamefont {B.}~\bibnamefont
  {Straughan}},\ }\bibfield  {title} {\bibinfo {title} {Thermal convection with
  the cattaneo–christov model},\ }\href@noop {} {\bibfield  {journal}
  {\bibinfo  {journal} {International Journal of Heat and Mass Transfer}\
  }\textbf {\bibinfo {volume} {53}},\ \bibinfo {pages} {95} (\bibinfo {year}
  {2010})}\BibitemShut {NoStop}%
\bibitem [{\citenamefont {Bray}\ and\ \citenamefont
  {Launder}(1995)}]{doi:10.1098/rspa.1995.0124}%
  \BibitemOpen
  \bibfield  {author} {\bibinfo {author} {\bibfnamefont {K.~N.~C.}\
  \bibnamefont {Bray}}\ and\ \bibinfo {author} {\bibfnamefont {B.~E.}\
  \bibnamefont {Launder}},\ }\bibfield  {title} {\bibinfo {title} {Turbulent
  transport in flames},\ }\href {https://doi.org/10.1098/rspa.1995.0124}
  {\bibfield  {journal} {\bibinfo  {journal} {Proceedings of the Royal Society
  of London. Series A: Mathematical and Physical Sciences}\ }\textbf {\bibinfo
  {volume} {451}},\ \bibinfo {pages} {231} (\bibinfo {year}
  {1995})}\BibitemShut {NoStop}%
\bibitem [{\citenamefont {Berezovskaya}\ and\ \citenamefont
  {Karev}(1999)}]{Berezovskaya_1999}%
  \BibitemOpen
  \bibfield  {author} {\bibinfo {author} {\bibfnamefont {F.~S.}\ \bibnamefont
  {Berezovskaya}}\ and\ \bibinfo {author} {\bibfnamefont {G.~P.}\ \bibnamefont
  {Karev}},\ }\bibfield  {title} {\bibinfo {title} {Bifurcations of travelling
  waves in population taxis models},\ }\href
  {https://doi.org/10.1070/PU1999v042n09ABEH000564} {\bibfield  {journal}
  {\bibinfo  {journal} {Physics-Uspekhi}\ }\textbf {\bibinfo {volume} {42}},\
  \bibinfo {pages} {917} (\bibinfo {year} {1999})}\BibitemShut {NoStop}%
\bibitem [{\citenamefont {Jordan}(2005)}]{JORDAN2005220}%
  \BibitemOpen
  \bibfield  {author} {\bibinfo {author} {\bibfnamefont {P.}~\bibnamefont
  {Jordan}},\ }\bibfield  {title} {\bibinfo {title} {Growth and decay of shock
  and acceleration waves in a traffic flow model with relaxation},\ }\href
  {https://doi.org/https://doi.org/10.1016/j.physd.2005.06.002} {\bibfield
  {journal} {\bibinfo  {journal} {Physica D: Nonlinear Phenomena}\ }\textbf
  {\bibinfo {volume} {207}},\ \bibinfo {pages} {220} (\bibinfo {year}
  {2005})}\BibitemShut {NoStop}%
\bibitem [{\citenamefont {Lucas}\ and\ \citenamefont
  {Fong}(2018)}]{Lucas_2018}%
  \BibitemOpen
  \bibfield  {author} {\bibinfo {author} {\bibfnamefont {A.}~\bibnamefont
  {Lucas}}\ and\ \bibinfo {author} {\bibfnamefont {K.~C.}\ \bibnamefont
  {Fong}},\ }\bibfield  {title} {\bibinfo {title} {Hydrodynamics of electrons
  in graphene},\ }\href {https://doi.org/10.1088/1361-648x/aaa274} {\bibfield
  {journal} {\bibinfo  {journal} {Journal of Physics: Condensed Matter}\
  }\textbf {\bibinfo {volume} {30}},\ \bibinfo {pages} {053001} (\bibinfo
  {year} {2018})}\BibitemShut {NoStop}%
\bibitem [{\citenamefont {Tranquada}(2020)}]{Tranquada_2020}%
  \BibitemOpen
  \bibfield  {author} {\bibinfo {author} {\bibfnamefont {J.~M.}\ \bibnamefont
  {Tranquada}},\ }\bibfield  {title} {\bibinfo {title} {Cuprate superconductors
  as viewed through a striped lens},\ }\href
  {https://doi.org/10.1080/00018732.2021.1935698} {\bibfield  {journal}
  {\bibinfo  {journal} {Advances in Physics}\ }\textbf {\bibinfo {volume}
  {69}},\ \bibinfo {pages} {437–509} (\bibinfo {year} {2020})}\BibitemShut
  {NoStop}%
\bibitem [{\citenamefont {Chaikin}\ and\ \citenamefont
  {Lubensky}(2000)}]{chaikin2000principles}%
  \BibitemOpen
  \bibfield  {author} {\bibinfo {author} {\bibfnamefont {P.}~\bibnamefont
  {Chaikin}}\ and\ \bibinfo {author} {\bibfnamefont {T.}~\bibnamefont
  {Lubensky}},\ }\href@noop {} {\emph {\bibinfo {title} {Principles of
  Condensed Matter Physics}}}\ (\bibinfo  {publisher} {Cambridge University
  Press},\ \bibinfo {year} {2000})\BibitemShut {NoStop}%
\bibitem [{\citenamefont {Yu}\ \emph {et~al.}(2020)\citenamefont {Yu},
  \citenamefont {Awe}, \citenamefont {Cochrane}, \citenamefont {Yates},
  \citenamefont {Hutchinson}, \citenamefont {Peterson},\ and\ \citenamefont
  {Bauer}}]{10.1063/1.5143271}%
  \BibitemOpen
  \bibfield  {author} {\bibinfo {author} {\bibfnamefont {E.~P.}\ \bibnamefont
  {Yu}}, \bibinfo {author} {\bibfnamefont {T.~J.}\ \bibnamefont {Awe}},
  \bibinfo {author} {\bibfnamefont {K.~R.}\ \bibnamefont {Cochrane}}, \bibinfo
  {author} {\bibfnamefont {K.~C.}\ \bibnamefont {Yates}}, \bibinfo {author}
  {\bibfnamefont {T.~M.}\ \bibnamefont {Hutchinson}}, \bibinfo {author}
  {\bibfnamefont {K.~J.}\ \bibnamefont {Peterson}},\ and\ \bibinfo {author}
  {\bibfnamefont {B.~S.}\ \bibnamefont {Bauer}},\ }\bibfield  {title} {\bibinfo
  {title} {{Use of hydrodynamic theory to estimate electrical current
  redistribution in metals}},\ }\href {https://doi.org/10.1063/1.5143271}
  {\bibfield  {journal} {\bibinfo  {journal} {Physics of Plasmas}\ }\textbf
  {\bibinfo {volume} {27}},\ \bibinfo {pages} {052703} (\bibinfo {year}
  {2020})}\BibitemShut {NoStop}%
\bibitem [{\citenamefont {Chen}\ \emph {et~al.}(2023)\citenamefont {Chen},
  \citenamefont {Lowder}, \citenamefont {Bakali}, \citenamefont {Andrews},
  \citenamefont {Schrenk}, \citenamefont {Waas}, \citenamefont {Svagera},
  \citenamefont {Eguchi}, \citenamefont {Prochaska}, \citenamefont {Wang},
  \citenamefont {Setty}, \citenamefont {Sur}, \citenamefont {Si}, \citenamefont
  {Paschen},\ and\ \citenamefont {Natelson}}]{doi:10.1126/science.abq6100}%
  \BibitemOpen
  \bibfield  {author} {\bibinfo {author} {\bibfnamefont {L.}~\bibnamefont
  {Chen}}, \bibinfo {author} {\bibfnamefont {D.~T.}\ \bibnamefont {Lowder}},
  \bibinfo {author} {\bibfnamefont {E.}~\bibnamefont {Bakali}}, \bibinfo
  {author} {\bibfnamefont {A.~M.}\ \bibnamefont {Andrews}}, \bibinfo {author}
  {\bibfnamefont {W.}~\bibnamefont {Schrenk}}, \bibinfo {author} {\bibfnamefont
  {M.}~\bibnamefont {Waas}}, \bibinfo {author} {\bibfnamefont {R.}~\bibnamefont
  {Svagera}}, \bibinfo {author} {\bibfnamefont {G.}~\bibnamefont {Eguchi}},
  \bibinfo {author} {\bibfnamefont {L.}~\bibnamefont {Prochaska}}, \bibinfo
  {author} {\bibfnamefont {Y.}~\bibnamefont {Wang}}, \bibinfo {author}
  {\bibfnamefont {C.}~\bibnamefont {Setty}}, \bibinfo {author} {\bibfnamefont
  {S.}~\bibnamefont {Sur}}, \bibinfo {author} {\bibfnamefont {Q.}~\bibnamefont
  {Si}}, \bibinfo {author} {\bibfnamefont {S.}~\bibnamefont {Paschen}},\ and\
  \bibinfo {author} {\bibfnamefont {D.}~\bibnamefont {Natelson}},\ }\bibfield
  {title} {\bibinfo {title} {Shot noise in a strange metal},\ }\href
  {https://doi.org/10.1126/science.abq6100} {\bibfield  {journal} {\bibinfo
  {journal} {Science}\ }\textbf {\bibinfo {volume} {382}},\ \bibinfo {pages}
  {907} (\bibinfo {year} {2023})}\BibitemShut {NoStop}%
\bibitem [{\citenamefont {Kovtun}(2016)}]{Kovtun:2016lfw}%
  \BibitemOpen
  \bibfield  {author} {\bibinfo {author} {\bibfnamefont {P.}~\bibnamefont
  {Kovtun}},\ }\bibfield  {title} {\bibinfo {title} {{Thermodynamics of
  polarized relativistic matter}},\ }\href
  {https://doi.org/10.1007/JHEP07(2016)028} {\bibfield  {journal} {\bibinfo
  {journal} {JHEP}\ }\textbf {\bibinfo {volume} {07}},\ \bibinfo {pages}
  {028}},\ \Eprint {https://arxiv.org/abs/1606.01226} {arXiv:1606.01226
  [hep-th]} \BibitemShut {NoStop}%
\bibitem [{\citenamefont {de~Boer}\ \emph {et~al.}(2020)\citenamefont
  {de~Boer}, \citenamefont {Hartong}, \citenamefont {Have}, \citenamefont
  {Obers},\ and\ \citenamefont {Sybesma}}]{deBoer:2020xlc}%
  \BibitemOpen
  \bibfield  {author} {\bibinfo {author} {\bibfnamefont {J.}~\bibnamefont
  {de~Boer}}, \bibinfo {author} {\bibfnamefont {J.}~\bibnamefont {Hartong}},
  \bibinfo {author} {\bibfnamefont {E.}~\bibnamefont {Have}}, \bibinfo {author}
  {\bibfnamefont {N.~A.}\ \bibnamefont {Obers}},\ and\ \bibinfo {author}
  {\bibfnamefont {W.}~\bibnamefont {Sybesma}},\ }\bibfield  {title} {\bibinfo
  {title} {{Non-Boost Invariant Fluid Dynamics}},\ }\href
  {https://doi.org/10.21468/SciPostPhys.9.2.018} {\bibfield  {journal}
  {\bibinfo  {journal} {SciPost Phys.}\ }\textbf {\bibinfo {volume} {9}},\
  \bibinfo {pages} {018} (\bibinfo {year} {2020})},\ \Eprint
  {https://arxiv.org/abs/2004.10759} {arXiv:2004.10759 [hep-th]} \BibitemShut
  {NoStop}%
\bibitem [{\citenamefont {Amoretti}\ \emph
  {et~al.}(2023{\natexlab{a}})\citenamefont {Amoretti}, \citenamefont
  {Brattan}, \citenamefont {Martinoia},\ and\ \citenamefont
  {Matthaiakakis}}]{Amoretti:2022ovc}%
  \BibitemOpen
  \bibfield  {author} {\bibinfo {author} {\bibfnamefont {A.}~\bibnamefont
  {Amoretti}}, \bibinfo {author} {\bibfnamefont {D.~K.}\ \bibnamefont
  {Brattan}}, \bibinfo {author} {\bibfnamefont {L.}~\bibnamefont {Martinoia}},\
  and\ \bibinfo {author} {\bibfnamefont {I.}~\bibnamefont {Matthaiakakis}},\
  }\bibfield  {title} {\bibinfo {title} {{Non-dissipative electrically driven
  fluids}},\ }\href {https://doi.org/10.1007/JHEP05(2023)218} {\bibfield
  {journal} {\bibinfo  {journal} {JHEP}\ }\textbf {\bibinfo {volume} {05}},\
  \bibinfo {pages} {218}},\ \Eprint {https://arxiv.org/abs/2211.05791}
  {arXiv:2211.05791 [hep-th]} \BibitemShut {NoStop}%
\bibitem [{\citenamefont {Amoretti}\ \emph {et~al.}(2024)\citenamefont
  {Amoretti}, \citenamefont {Brattan}, \citenamefont {Martinoia},\ and\
  \citenamefont {Rongen}}]{Amoretti:2024jig}%
  \BibitemOpen
  \bibfield  {author} {\bibinfo {author} {\bibfnamefont {A.}~\bibnamefont
  {Amoretti}}, \bibinfo {author} {\bibfnamefont {D.~K.}\ \bibnamefont
  {Brattan}}, \bibinfo {author} {\bibfnamefont {L.}~\bibnamefont {Martinoia}},\
  and\ \bibinfo {author} {\bibfnamefont {J.}~\bibnamefont {Rongen}},\
  }\bibfield  {title} {\bibinfo {title} {Dissipative electrically driven
  fluids},\ }\href@noop {} {\bibfield  {journal} {\bibinfo  {journal} {arXiv
  preprint arXiv:2407.18856}\ } (\bibinfo {year} {2024})}\BibitemShut {NoStop}%
\bibitem [{\citenamefont {Karch}\ and\ \citenamefont
  {Katz}(2002)}]{Karch:2002sh}%
  \BibitemOpen
  \bibfield  {author} {\bibinfo {author} {\bibfnamefont {A.}~\bibnamefont
  {Karch}}\ and\ \bibinfo {author} {\bibfnamefont {E.}~\bibnamefont {Katz}},\
  }\bibfield  {title} {\bibinfo {title} {{Adding flavor to AdS / CFT}},\ }\href
  {https://doi.org/10.1088/1126-6708/2002/06/043} {\bibfield  {journal}
  {\bibinfo  {journal} {JHEP}\ }\textbf {\bibinfo {volume} {06}},\ \bibinfo
  {pages} {043}},\ \Eprint {https://arxiv.org/abs/hep-th/0205236}
  {arXiv:hep-th/0205236} \BibitemShut {NoStop}%
\bibitem [{\citenamefont {Baggioli}\ \emph {et~al.}(2023)\citenamefont
  {Baggioli}, \citenamefont {Bu},\ and\ \citenamefont
  {Ziogas}}]{Baggioli:2023tlc}%
  \BibitemOpen
  \bibfield  {author} {\bibinfo {author} {\bibfnamefont {M.}~\bibnamefont
  {Baggioli}}, \bibinfo {author} {\bibfnamefont {Y.}~\bibnamefont {Bu}},\ and\
  \bibinfo {author} {\bibfnamefont {V.}~\bibnamefont {Ziogas}},\ }\bibfield
  {title} {\bibinfo {title} {{U(1) quasi-hydrodynamics: Schwinger-Keldysh
  effective field theory and holography}},\ }\href
  {https://doi.org/10.1007/JHEP09(2023)019} {\bibfield  {journal} {\bibinfo
  {journal} {JHEP}\ }\textbf {\bibinfo {volume} {09}},\ \bibinfo {pages}
  {019}},\ \Eprint {https://arxiv.org/abs/2304.14173} {arXiv:2304.14173
  [hep-th]} \BibitemShut {NoStop}%
\bibitem [{\citenamefont {Amoretti}\ \emph
  {et~al.}(2023{\natexlab{b}})\citenamefont {Amoretti}, \citenamefont
  {Brattan}, \citenamefont {Martinoia},\ and\ \citenamefont
  {Matthaiakakis}}]{Amoretti:2023vhe}%
  \BibitemOpen
  \bibfield  {author} {\bibinfo {author} {\bibfnamefont {A.}~\bibnamefont
  {Amoretti}}, \bibinfo {author} {\bibfnamefont {D.~K.}\ \bibnamefont
  {Brattan}}, \bibinfo {author} {\bibfnamefont {L.}~\bibnamefont {Martinoia}},\
  and\ \bibinfo {author} {\bibfnamefont {I.}~\bibnamefont {Matthaiakakis}},\
  }\bibfield  {title} {\bibinfo {title} {{Restoring time-reversal covariance in
  relaxed hydrodynamics}},\ }\href
  {https://doi.org/10.1103/PhysRevD.108.056003} {\bibfield  {journal} {\bibinfo
   {journal} {Phys. Rev. D}\ }\textbf {\bibinfo {volume} {108}},\ \bibinfo
  {pages} {056003} (\bibinfo {year} {2023}{\natexlab{b}})},\ \Eprint
  {https://arxiv.org/abs/2304.01248} {arXiv:2304.01248 [hep-th]} \BibitemShut
  {NoStop}%
\bibitem [{\citenamefont {Chen}\ and\ \citenamefont
  {Lucas}(2017)}]{Chen:2017dsy}%
  \BibitemOpen
  \bibfield  {author} {\bibinfo {author} {\bibfnamefont {C.-F.}\ \bibnamefont
  {Chen}}\ and\ \bibinfo {author} {\bibfnamefont {A.}~\bibnamefont {Lucas}},\
  }\bibfield  {title} {\bibinfo {title} {{Origin of the Drude peak and of zero
  sound in probe brane holography}},\ }\href
  {https://doi.org/10.1016/j.physletb.2017.10.023} {\bibfield  {journal}
  {\bibinfo  {journal} {Phys. Lett. B}\ }\textbf {\bibinfo {volume} {774}},\
  \bibinfo {pages} {569} (\bibinfo {year} {2017})},\ \Eprint
  {https://arxiv.org/abs/1709.01520} {arXiv:1709.01520 [hep-th]} \BibitemShut
  {NoStop}%
\bibitem [{\citenamefont {Kovtun}(2012)}]{Kovtun:2012rj}%
  \BibitemOpen
  \bibfield  {author} {\bibinfo {author} {\bibfnamefont {P.}~\bibnamefont
  {Kovtun}},\ }\bibfield  {title} {\bibinfo {title} {{Lectures on hydrodynamic
  fluctuations in relativistic theories}},\ }\href
  {https://doi.org/10.1088/1751-8113/45/47/473001} {\bibfield  {journal}
  {\bibinfo  {journal} {J. Phys. A}\ }\textbf {\bibinfo {volume} {45}},\
  \bibinfo {pages} {473001} (\bibinfo {year} {2012})},\ \Eprint
  {https://arxiv.org/abs/1205.5040} {arXiv:1205.5040 [hep-th]} \BibitemShut
  {NoStop}%
\bibitem [{\citenamefont {Ammon}\ and\ \citenamefont
  {Erdmenger}(2015)}]{Ammon:2015wua}%
  \BibitemOpen
  \bibfield  {author} {\bibinfo {author} {\bibfnamefont {M.}~\bibnamefont
  {Ammon}}\ and\ \bibinfo {author} {\bibfnamefont {J.}~\bibnamefont
  {Erdmenger}},\ }\href {https://doi.org/10.1017/CBO9780511846373} {\emph
  {\bibinfo {title} {{Gauge/gravity duality}: {Foundations and
  applications}}}}\ (\bibinfo  {publisher} {Cambridge University Press},\
  \bibinfo {address} {Cambridge},\ \bibinfo {year} {2015})\BibitemShut
  {NoStop}%
\bibitem [{\citenamefont {Liu}\ and\ \citenamefont {Sonner}(2019)}]{Liu_2020}%
  \BibitemOpen
  \bibfield  {author} {\bibinfo {author} {\bibfnamefont {H.}~\bibnamefont
  {Liu}}\ and\ \bibinfo {author} {\bibfnamefont {J.}~\bibnamefont {Sonner}},\
  }\bibfield  {title} {\bibinfo {title} {Holographic systems far from
  equilibrium: a review},\ }\href {https://doi.org/10.1088/1361-6633/ab4f91}
  {\bibfield  {journal} {\bibinfo  {journal} {Reports on Progress in Physics}\
  }\textbf {\bibinfo {volume} {83}},\ \bibinfo {pages} {016001} (\bibinfo
  {year} {2019})}\BibitemShut {NoStop}%
\bibitem [{\citenamefont {Karch}\ and\ \citenamefont
  {O'Bannon}(2007)}]{Karch:2007pd}%
  \BibitemOpen
  \bibfield  {author} {\bibinfo {author} {\bibfnamefont {A.}~\bibnamefont
  {Karch}}\ and\ \bibinfo {author} {\bibfnamefont {A.}~\bibnamefont
  {O'Bannon}},\ }\bibfield  {title} {\bibinfo {title} {{Metallic AdS/CFT}},\
  }\href {https://doi.org/10.1088/1126-6708/2007/09/024} {\bibfield  {journal}
  {\bibinfo  {journal} {JHEP}\ }\textbf {\bibinfo {volume} {09}},\ \bibinfo
  {pages} {024}},\ \Eprint {https://arxiv.org/abs/0705.3870} {arXiv:0705.3870
  [hep-th]} \BibitemShut {NoStop}%
\bibitem [{\citenamefont {Son}\ and\ \citenamefont
  {Starinets}(2002)}]{Son:2002sd}%
  \BibitemOpen
  \bibfield  {author} {\bibinfo {author} {\bibfnamefont {D.~T.}\ \bibnamefont
  {Son}}\ and\ \bibinfo {author} {\bibfnamefont {A.~O.}\ \bibnamefont
  {Starinets}},\ }\bibfield  {title} {\bibinfo {title} {{Minkowski space
  correlators in AdS / CFT correspondence: Recipe and applications}},\ }\href
  {https://doi.org/10.1088/1126-6708/2002/09/042} {\bibfield  {journal}
  {\bibinfo  {journal} {JHEP}\ }\textbf {\bibinfo {volume} {09}},\ \bibinfo
  {pages} {042}},\ \Eprint {https://arxiv.org/abs/hep-th/0205051}
  {arXiv:hep-th/0205051} \BibitemShut {NoStop}%
\bibitem [{\citenamefont {Mas}\ \emph {et~al.}(2009)\citenamefont {Mas},
  \citenamefont {Shock},\ and\ \citenamefont {Tarrio}}]{Mas:2009wf}%
  \BibitemOpen
  \bibfield  {author} {\bibinfo {author} {\bibfnamefont {J.}~\bibnamefont
  {Mas}}, \bibinfo {author} {\bibfnamefont {J.~P.}\ \bibnamefont {Shock}},\
  and\ \bibinfo {author} {\bibfnamefont {J.}~\bibnamefont {Tarrio}},\
  }\bibfield  {title} {\bibinfo {title} {{Holographic Spectral Functions in
  Metallic AdS/CFT}},\ }\href {https://doi.org/10.1088/1126-6708/2009/09/032}
  {\bibfield  {journal} {\bibinfo  {journal} {JHEP}\ }\textbf {\bibinfo
  {volume} {09}},\ \bibinfo {pages} {032}},\ \Eprint
  {https://arxiv.org/abs/0904.3905} {arXiv:0904.3905 [hep-th]} \BibitemShut
  {NoStop}%
\bibitem [{\citenamefont {Kaminski}\ \emph {et~al.}(2010)\citenamefont
  {Kaminski}, \citenamefont {Landsteiner}, \citenamefont {Mas}, \citenamefont
  {Shock},\ and\ \citenamefont {Tarrio}}]{Kaminski:2009dh}%
  \BibitemOpen
  \bibfield  {author} {\bibinfo {author} {\bibfnamefont {M.}~\bibnamefont
  {Kaminski}}, \bibinfo {author} {\bibfnamefont {K.}~\bibnamefont
  {Landsteiner}}, \bibinfo {author} {\bibfnamefont {J.}~\bibnamefont {Mas}},
  \bibinfo {author} {\bibfnamefont {J.~P.}\ \bibnamefont {Shock}},\ and\
  \bibinfo {author} {\bibfnamefont {J.}~\bibnamefont {Tarrio}},\ }\bibfield
  {title} {\bibinfo {title} {{Holographic Operator Mixing and Quasinormal Modes
  on the Brane}},\ }\href {https://doi.org/10.1007/JHEP02(2010)021} {\bibfield
  {journal} {\bibinfo  {journal} {JHEP}\ }\textbf {\bibinfo {volume} {02}},\
  \bibinfo {pages} {021}},\ \Eprint {https://arxiv.org/abs/0911.3610}
  {arXiv:0911.3610 [hep-th]} \BibitemShut {NoStop}%
\bibitem [{\citenamefont {Sonner}\ and\ \citenamefont
  {Green}(2012)}]{Sonner:2012if}%
  \BibitemOpen
  \bibfield  {author} {\bibinfo {author} {\bibfnamefont {J.}~\bibnamefont
  {Sonner}}\ and\ \bibinfo {author} {\bibfnamefont {A.~G.}\ \bibnamefont
  {Green}},\ }\bibfield  {title} {\bibinfo {title} {{Hawking Radiation and
  Non-equilibrium Quantum Critical Current Noise}},\ }\href
  {https://doi.org/10.1103/PhysRevLett.109.091601} {\bibfield  {journal}
  {\bibinfo  {journal} {Phys. Rev. Lett.}\ }\textbf {\bibinfo {volume} {109}},\
  \bibinfo {pages} {091601} (\bibinfo {year} {2012})},\ \Eprint
  {https://arxiv.org/abs/1203.4908} {arXiv:1203.4908 [cond-mat.str-el]}
  \BibitemShut {NoStop}%
\bibitem [{\citenamefont {Sulpizio}\ \emph {et~al.}(2019)\citenamefont
  {Sulpizio}, \citenamefont {Ella}, \citenamefont {Rozen}, \citenamefont
  {Birkbeck}, \citenamefont {Perello}, \citenamefont {Dutta}, \citenamefont
  {Ben-Shalom}, \citenamefont {Taniguchi}, \citenamefont {Watanabe},
  \citenamefont {Holder}, \citenamefont {Queiroz}, \citenamefont {Principi},
  \citenamefont {Stern}, \citenamefont {Scaffidi}, \citenamefont {Geim},\ and\
  \citenamefont {Ilani}}]{Sulpizio2019}%
  \BibitemOpen
  \bibfield  {author} {\bibinfo {author} {\bibfnamefont {J.~A.}\ \bibnamefont
  {Sulpizio}}, \bibinfo {author} {\bibfnamefont {L.}~\bibnamefont {Ella}},
  \bibinfo {author} {\bibfnamefont {A.}~\bibnamefont {Rozen}}, \bibinfo
  {author} {\bibfnamefont {J.}~\bibnamefont {Birkbeck}}, \bibinfo {author}
  {\bibfnamefont {D.~J.}\ \bibnamefont {Perello}}, \bibinfo {author}
  {\bibfnamefont {D.}~\bibnamefont {Dutta}}, \bibinfo {author} {\bibfnamefont
  {M.}~\bibnamefont {Ben-Shalom}}, \bibinfo {author} {\bibfnamefont
  {T.}~\bibnamefont {Taniguchi}}, \bibinfo {author} {\bibfnamefont
  {K.}~\bibnamefont {Watanabe}}, \bibinfo {author} {\bibfnamefont
  {T.}~\bibnamefont {Holder}}, \bibinfo {author} {\bibfnamefont
  {R.}~\bibnamefont {Queiroz}}, \bibinfo {author} {\bibfnamefont
  {A.}~\bibnamefont {Principi}}, \bibinfo {author} {\bibfnamefont
  {A.}~\bibnamefont {Stern}}, \bibinfo {author} {\bibfnamefont
  {T.}~\bibnamefont {Scaffidi}}, \bibinfo {author} {\bibfnamefont {A.~K.}\
  \bibnamefont {Geim}},\ and\ \bibinfo {author} {\bibfnamefont
  {S.}~\bibnamefont {Ilani}},\ }\bibfield  {title} {\bibinfo {title}
  {Visualizing poiseuille flow of hydrodynamic electrons},\ }\href
  {https://doi.org/10.1038/s41586-019-1788-9} {\bibfield  {journal} {\bibinfo
  {journal} {Nature}\ }\textbf {\bibinfo {volume} {576}},\ \bibinfo {pages}
  {75} (\bibinfo {year} {2019})}\BibitemShut {NoStop}%
\bibitem [{\citenamefont {Ku}\ \emph {et~al.}(2020)\citenamefont {Ku},
  \citenamefont {Zhou}, \citenamefont {Li}, \citenamefont {Shin}, \citenamefont
  {Shi}, \citenamefont {Burch}, \citenamefont {Anderson}, \citenamefont
  {Pierce}, \citenamefont {Xie}, \citenamefont {Hamo}, \citenamefont {Vool},
  \citenamefont {Zhang}, \citenamefont {Casola}, \citenamefont {Taniguchi},
  \citenamefont {Watanabe}, \citenamefont {Fogler}, \citenamefont {Kim},
  \citenamefont {Yacoby},\ and\ \citenamefont {Walsworth}}]{Ku2020}%
  \BibitemOpen
  \bibfield  {author} {\bibinfo {author} {\bibfnamefont {M.~J.~H.}\
  \bibnamefont {Ku}}, \bibinfo {author} {\bibfnamefont {T.~X.}\ \bibnamefont
  {Zhou}}, \bibinfo {author} {\bibfnamefont {Q.}~\bibnamefont {Li}}, \bibinfo
  {author} {\bibfnamefont {Y.~J.}\ \bibnamefont {Shin}}, \bibinfo {author}
  {\bibfnamefont {J.~K.}\ \bibnamefont {Shi}}, \bibinfo {author} {\bibfnamefont
  {C.}~\bibnamefont {Burch}}, \bibinfo {author} {\bibfnamefont {L.~E.}\
  \bibnamefont {Anderson}}, \bibinfo {author} {\bibfnamefont {A.~T.}\
  \bibnamefont {Pierce}}, \bibinfo {author} {\bibfnamefont {Y.}~\bibnamefont
  {Xie}}, \bibinfo {author} {\bibfnamefont {A.}~\bibnamefont {Hamo}}, \bibinfo
  {author} {\bibfnamefont {U.}~\bibnamefont {Vool}}, \bibinfo {author}
  {\bibfnamefont {H.}~\bibnamefont {Zhang}}, \bibinfo {author} {\bibfnamefont
  {F.}~\bibnamefont {Casola}}, \bibinfo {author} {\bibfnamefont
  {T.}~\bibnamefont {Taniguchi}}, \bibinfo {author} {\bibfnamefont
  {K.}~\bibnamefont {Watanabe}}, \bibinfo {author} {\bibfnamefont {M.~M.}\
  \bibnamefont {Fogler}}, \bibinfo {author} {\bibfnamefont {P.}~\bibnamefont
  {Kim}}, \bibinfo {author} {\bibfnamefont {A.}~\bibnamefont {Yacoby}},\ and\
  \bibinfo {author} {\bibfnamefont {R.~L.}\ \bibnamefont {Walsworth}},\
  }\bibfield  {title} {\bibinfo {title} {Imaging viscous flow of the dirac
  fluid in graphene},\ }\href {https://doi.org/10.1038/s41586-020-2507-2}
  {\bibfield  {journal} {\bibinfo  {journal} {Nature}\ }\textbf {\bibinfo
  {volume} {583}},\ \bibinfo {pages} {537} (\bibinfo {year}
  {2020})}\BibitemShut {NoStop}%
\bibitem [{\citenamefont {Mitrano}\ \emph {et~al.}(2019)\citenamefont
  {Mitrano}, \citenamefont {Lee}, \citenamefont {Husain}, \citenamefont
  {Delacretaz}, \citenamefont {Zhu}, \citenamefont {de~la Peña~Munoz},
  \citenamefont {Sun}, \citenamefont {Joe}, \citenamefont {Reid}, \citenamefont
  {Wandel}, \citenamefont {Coslovich}, \citenamefont {Schlotter}, \citenamefont
  {van Driel}, \citenamefont {Schneeloch}, \citenamefont {Gu}, \citenamefont
  {Hartnoll}, \citenamefont {Goldenfeld},\ and\ \citenamefont
  {Abbamonte}}]{doi:10.1126/sciadv.aax3346}%
  \BibitemOpen
  \bibfield  {author} {\bibinfo {author} {\bibfnamefont {M.}~\bibnamefont
  {Mitrano}}, \bibinfo {author} {\bibfnamefont {S.}~\bibnamefont {Lee}},
  \bibinfo {author} {\bibfnamefont {A.~A.}\ \bibnamefont {Husain}}, \bibinfo
  {author} {\bibfnamefont {L.}~\bibnamefont {Delacretaz}}, \bibinfo {author}
  {\bibfnamefont {M.}~\bibnamefont {Zhu}}, \bibinfo {author} {\bibfnamefont
  {G.}~\bibnamefont {de~la Peña~Munoz}}, \bibinfo {author} {\bibfnamefont
  {S.~X.-L.}\ \bibnamefont {Sun}}, \bibinfo {author} {\bibfnamefont {Y.~I.}\
  \bibnamefont {Joe}}, \bibinfo {author} {\bibfnamefont {A.~H.}\ \bibnamefont
  {Reid}}, \bibinfo {author} {\bibfnamefont {S.~F.}\ \bibnamefont {Wandel}},
  \bibinfo {author} {\bibfnamefont {G.}~\bibnamefont {Coslovich}}, \bibinfo
  {author} {\bibfnamefont {W.}~\bibnamefont {Schlotter}}, \bibinfo {author}
  {\bibfnamefont {T.}~\bibnamefont {van Driel}}, \bibinfo {author}
  {\bibfnamefont {J.}~\bibnamefont {Schneeloch}}, \bibinfo {author}
  {\bibfnamefont {G.~D.}\ \bibnamefont {Gu}}, \bibinfo {author} {\bibfnamefont
  {S.}~\bibnamefont {Hartnoll}}, \bibinfo {author} {\bibfnamefont
  {N.}~\bibnamefont {Goldenfeld}},\ and\ \bibinfo {author} {\bibfnamefont
  {P.}~\bibnamefont {Abbamonte}},\ }\bibfield  {title} {\bibinfo {title}
  {Ultrafast time-resolved x-ray scattering reveals diffusive charge order
  dynamics in {La2$-$xBaxCuO4}},\ }\href
  {https://doi.org/10.1126/sciadv.aax3346} {\bibfield  {journal} {\bibinfo
  {journal} {Science Advances}\ }\textbf {\bibinfo {volume} {5}},\ \bibinfo
  {pages} {eaax3346} (\bibinfo {year} {2019})}\BibitemShut {NoStop}%
\bibitem [{\citenamefont {Amoretti}\ and\ \citenamefont
  {Brattan}(2022)}]{Amoretti:2022acb}%
  \BibitemOpen
  \bibfield  {author} {\bibinfo {author} {\bibfnamefont {A.}~\bibnamefont
  {Amoretti}}\ and\ \bibinfo {author} {\bibfnamefont {D.~K.}\ \bibnamefont
  {Brattan}},\ }\bibfield  {title} {\bibinfo {title} {{On the hydrodynamics of
  (2 + 1)-dimensional strongly coupled relativistic theories in an external
  magnetic field}},\ }\href {https://doi.org/10.1142/S0217732322300105}
  {\bibfield  {journal} {\bibinfo  {journal} {Mod. Phys. Lett. A}\ }\textbf
  {\bibinfo {volume} {37}},\ \bibinfo {pages} {2230010} (\bibinfo {year}
  {2022})},\ \Eprint {https://arxiv.org/abs/2209.11589} {arXiv:2209.11589
  [hep-th]} \BibitemShut {NoStop}%
\bibitem [{\citenamefont {Amoretti}\ \emph
  {et~al.}(2023{\natexlab{c}})\citenamefont {Amoretti}, \citenamefont
  {Brattan}, \citenamefont {Martinoia},\ and\ \citenamefont
  {Matthaiakakis}}]{Amoretti:2022vxq}%
  \BibitemOpen
  \bibfield  {author} {\bibinfo {author} {\bibfnamefont {A.}~\bibnamefont
  {Amoretti}}, \bibinfo {author} {\bibfnamefont {D.~K.}\ \bibnamefont
  {Brattan}}, \bibinfo {author} {\bibfnamefont {L.}~\bibnamefont {Martinoia}},\
  and\ \bibinfo {author} {\bibfnamefont {I.}~\bibnamefont {Matthaiakakis}},\
  }\bibfield  {title} {\bibinfo {title} {{Leading order magnetic field
  dependence of conductivities in anomalous hydrodynamics}},\ }\href
  {https://doi.org/10.1103/PhysRevD.108.016003} {\bibfield  {journal} {\bibinfo
   {journal} {Phys. Rev. D}\ }\textbf {\bibinfo {volume} {108}},\ \bibinfo
  {pages} {016003} (\bibinfo {year} {2023}{\natexlab{c}})},\ \Eprint
  {https://arxiv.org/abs/2212.09761} {arXiv:2212.09761 [hep-th]} \BibitemShut
  {NoStop}%
\bibitem [{\citenamefont {Brattan}\ \emph
  {et~al.}(2018{\natexlab{a}})\citenamefont {Brattan}, \citenamefont {Ovdat},\
  and\ \citenamefont {Akkermans}}]{Brattan:2017yzx}%
  \BibitemOpen
  \bibfield  {author} {\bibinfo {author} {\bibfnamefont {D.~K.}\ \bibnamefont
  {Brattan}}, \bibinfo {author} {\bibfnamefont {O.}~\bibnamefont {Ovdat}},\
  and\ \bibinfo {author} {\bibfnamefont {E.}~\bibnamefont {Akkermans}},\
  }\bibfield  {title} {\bibinfo {title} {{Scale anomaly of a Lifshitz scalar: a
  universal quantum phase transition to discrete scale invariance}},\ }\href
  {https://doi.org/10.1103/PhysRevD.97.061701} {\bibfield  {journal} {\bibinfo
  {journal} {Phys. Rev. D}\ }\textbf {\bibinfo {volume} {97}},\ \bibinfo
  {pages} {061701} (\bibinfo {year} {2018}{\natexlab{a}})},\ \Eprint
  {https://arxiv.org/abs/1706.00016} {arXiv:1706.00016 [hep-th]} \BibitemShut
  {NoStop}%
\bibitem [{\citenamefont {Brattan}(2018)}]{Brattan:2018sgc}%
  \BibitemOpen
  \bibfield  {author} {\bibinfo {author} {\bibfnamefont {D.~K.}\ \bibnamefont
  {Brattan}},\ }\bibfield  {title} {\bibinfo {title} {{$\mathcal{N}=2$
  supersymmetry and anisotropic scale invariance}},\ }\href
  {https://doi.org/10.1103/PhysRevD.98.036005} {\bibfield  {journal} {\bibinfo
  {journal} {Phys. Rev. D}\ }\textbf {\bibinfo {volume} {98}},\ \bibinfo
  {pages} {036005} (\bibinfo {year} {2018})},\ \Eprint
  {https://arxiv.org/abs/1801.03098} {arXiv:1801.03098 [hep-th]} \BibitemShut
  {NoStop}%
\bibitem [{\citenamefont {Brattan}\ \emph
  {et~al.}(2018{\natexlab{b}})\citenamefont {Brattan}, \citenamefont {Ovdat},\
  and\ \citenamefont {Akkermans}}]{Brattan:2018cgk}%
  \BibitemOpen
  \bibfield  {author} {\bibinfo {author} {\bibfnamefont {D.~K.}\ \bibnamefont
  {Brattan}}, \bibinfo {author} {\bibfnamefont {O.}~\bibnamefont {Ovdat}},\
  and\ \bibinfo {author} {\bibfnamefont {E.}~\bibnamefont {Akkermans}},\
  }\bibfield  {title} {\bibinfo {title} {{On the landscape of scale invariance
  in quantum mechanics}},\ }\href {https://doi.org/10.1088/1751-8121/aadfae}
  {\bibfield  {journal} {\bibinfo  {journal} {J. Phys. A}\ }\textbf {\bibinfo
  {volume} {51}},\ \bibinfo {pages} {435401} (\bibinfo {year}
  {2018}{\natexlab{b}})},\ \Eprint {https://arxiv.org/abs/1804.10213}
  {arXiv:1804.10213 [hep-th]} \BibitemShut {NoStop}%
\bibitem [{\citenamefont {Ishigaki}\ \emph {et~al.}(2022)\citenamefont
  {Ishigaki}, \citenamefont {Kinoshita},\ and\ \citenamefont
  {Matsumoto}}]{Ishigaki:2021vyv}%
  \BibitemOpen
  \bibfield  {author} {\bibinfo {author} {\bibfnamefont {S.}~\bibnamefont
  {Ishigaki}}, \bibinfo {author} {\bibfnamefont {S.}~\bibnamefont
  {Kinoshita}},\ and\ \bibinfo {author} {\bibfnamefont {M.}~\bibnamefont
  {Matsumoto}},\ }\bibfield  {title} {\bibinfo {title} {{Dynamical stability
  and filamentary instability in holographic conductors}},\ }\href
  {https://doi.org/10.1007/JHEP04(2022)173} {\bibfield  {journal} {\bibinfo
  {journal} {JHEP}\ }\textbf {\bibinfo {volume} {04}},\ \bibinfo {pages}
  {173}},\ \Eprint {https://arxiv.org/abs/2112.11677} {arXiv:2112.11677
  [hep-th]} \BibitemShut {NoStop}%
\end{thebibliography}
\end{document}